\documentclass[twocolumn,showpacs,amsmath,amssymb,pre]{revtex4}

\usepackage{graphicx}
\usepackage{dcolumn}
\usepackage{bm}

\begin{document}

\title{Relaxation in a glassy binary mixture: Comparison of the mode-coupling 
theory to a Brownian dynamics simulation}

\author{Elijah Flenner}
\author{Grzegorz Szamel}
\affiliation{Department of Chemistry, Colorado State University, Fort Collins, CO 80525}

\begin{abstract}
We solved the mode-coupling equations for the Kob-Andersen
binary mixture using the
structure factors calculated from Brownian dynamics
simulations of the same system.
We found, as was previously observed, that the mode-coupling temperature, $T_c$,
inferred from simulations is about two times greater than that
predicted by the theory.  However, we find that many time dependent
quantities agree reasonably well with the predictions of the
mode-coupling theory if they are compared at the same reduced temperature
$\epsilon = (T-T_c)/T_c$, and if $\epsilon$ is not too small.
Specifically, the simulation results for the 
incoherent intermediate scattering function, the mean square displacement,
the relaxation time and the self-diffusion coefficient 
agree reasonably well with the predictions of the mode-coupling theory.
We find that there are substantial differences for
the non-Gaussian parameter. At small reduced temperatures  
the probabilities of the logarithm of
single particle displacements demonstrate that there is hopping-like
motion present in the simulations, and this motion is not predicted by
the mode-coupling theory.  The wave vector dependent relaxation
time is shown to be qualitatively different than the predictions of the
mode-coupling theory for temperatures where hopping-like motion is present.
\end{abstract}

\date{\today}


\maketitle

\section{\label{intro}Introduction}

The mode-coupling theory \cite{Goetze,Das} has been used extensively to
describe the slow relaxation observed in supercooled liquids 
close to the glass transition. 
The qualitative predictions of the theory have been
compared to experiments \cite{Das,Lunkenheimer} and simulations 
\cite{Barrat,KobAndersen,Fuchs,Franosch}. 
However, until recently, quantitative comparisons which include
wave vector dependence and use exact (\textit{i.e.}~obtained from simulations) 
static information have
been rare \cite{Nauroth,KobNauroth,Foffi}.

Nauroth and Kob \cite{Nauroth} compared the non-ergodicity parameters 
predicted by the mode-coupling theory to
Newtonian dynamics simulations of a binary Lennard-Jones mixture.
They used the structure factors determined
from the simulations as the input to the mode-coupling theory,
and found that the transition temperature predicted
by the theory was approximately twice higher 
than the transition temperature inferred from simulations.  
However, the non-ergodicity parameters predicted by
the theory agreed reasonably well with the
ones determined from simulations.  
In a later work, Kob, Nauroth, and Sciortino \cite{KobNauroth} 
quantitatively compared the shape of the intermediate scattering functions predicted by
the mode-coupling theory to results of Newtonian dynamics simulations.  
Since the transition temperature predicted by the
theory is greater than that 
inferred from simulations, the intermediate scattering
functions at the same temperature (and close
to the transition temperature) are vastly different. 
Therefore, the authors compared scattering functions predicted
by the theory at a higher temperature to scattering functions
obtained from simulations at a lower temperature. The two corresponding temperatures
were determined by the requirement that the relaxation time be
the same for one wave vector
around the first peak in partial structure factor for the larger particles.
Once the two corresponding temperatures where fixed, the authors 
showed that for a few wave vectors the shape of the incoherent intermediate scattering
function was well described by the 
mode-coupling theory. Also, Foffi \textit{et al.}~\cite{Foffi} compared simulation
results to the mode-coupling theory for mixtures of hard 
spheres. They found that, if the time scale is rescaled, the mode-coupling theory
accurately predicts the shape of the intermediate scattering functions in 
the $\alpha$ relaxation region.

Recently, the mode-coupling theory has been compared to molecular
and Brownian dynamics simulations of polydisperse spheres
with a strong repulsive core by 
Voigtmann, Puertes, and Fuchs \cite{Voigtmann}.  They used 
the Percus-Yevick theory to determine the structure factors
for the mode-coupling theory calculations. 
For the comparison of the theory with simulation,
they adjusted the packing fraction and allowed a ``dynamical'' 
length scale to vary slightly.  They found that after these adjustments
the intermediate scattering functions and the mean square displacement
predicted by the mode-coupling theory agreed well with the simulation results.

The goal of the work presented here is to compare the predictions of the mode
coupling theory to the results of Brownian dynamics simulations using the smallest
possible number of adjustable parameters. In other words, we would like to 
test the predictive power of the mode-coupling theory rather than demonstrate
that by making a number of parameters adjustable, we can reproduce the
simulation results very accurately. In view of the difference between
the mode-coupling transition temperature predicted by the theory and
inferred from simulations, a minimalistic approach is to compare
theoretical predictions and results of the simulations at 
the same reduced temperature $\epsilon = (T - T_c)/T_c$. 
Briefly, the result of our comparison is that many time
dependent quantities predicted by the theory agree reasonably well
with results of the simulations if $\epsilon$ is not too small. However,
there is significant disagreement between the theoretical predictions 
and the simulation results 
for the non-Gaussian parameter for all
but the highest reduced temperatures. 
Finally, at low reduced temperatures there is a hopping-like 
motion present in the simulations which 
is not predicted by the mode-coupling theory. 

One of the reasons for testing the mode coupling theory against Brownian
dynamics simulations is that the approximations involved in this theory applied to 
Brownian systems are somewhat less severe. The first approximation of the theory,
projecting stress fluctuations onto the subspace of the density products \cite{Goetze},
is exact for Brownian systems. However, the main and the most drastic approximation of the
mode coupling theory, the self-consistent factorization approximation, has to still be used.
The present study shows explicitly that it is the factorization approximation that is 
responsible for the failure of the mode coupling theory for low reduced temperatures.

The paper is organized as follows.  In section \ref{simulations} we
briefly describe the Brownian dynamics simulations.  In 
section \ref{mct} we describe the mode-coupling equations
which are appropriate for a binary mixture evolving with Brownian dynamics, and
briefly present the mode-coupling calculation (the method is described in detail in the appendix).
In section \ref{transT} we describe the method used to find
the transition temperature from the mode-coupling theory, and 
briefly discuss the non-ergodicity parameters.  We  
compare the intermediate scattering functions in section \ref{scatt} and the
mean square displacement in section \ref{msd}.  We compare the
non-Gaussian parameter 
in section \ref{nongauss}, and the probability distribution
of the logarithm of single particle displacements in  
section \ref{vanhove}.  We discuss the results in section
\ref{conclusion}.

\section{\label{simulations}Brownian Dynamics Simulations}

We simulated a system consisting of $N_A=800$ particles of type A and
$N_B=200$ particles of type B that was first considered by Kob and Andersen \cite{KobAndersen}.
The interaction potential is
$V_{\alpha \beta}(r) = 4\epsilon_{\alpha \beta}[
({\sigma_{\alpha \beta}}/{r})^{12} - ({\sigma_{\alpha \beta}}/{r})^6]$,
where $\alpha, \beta \in \{A,B\}$, $\epsilon_{AA} = 1.0$, 
$\sigma_{AA} = 1.0$, $\epsilon_{AB} = 1.5$, $\sigma_{AB} = 0.8$, 
$\epsilon_{BB} = 0.5$, and $\sigma_{BB} = 0.88$.  The simulations
are performed with the interaction potential cut at 
$2.5\ \sigma_{\alpha \beta}$, and the box length of the cubic simulation cell
is $9.4\ \sigma_{AA}$.  Periodic boundary conditions were used.

We performed Brownian dynamics simulations.  The equation of 
motion for the position of the $i_{th}$ particle of type $\alpha$, $\vec{r}\,_{i}^{\alpha}$, is
\begin{equation}\label{Lang}
\dot{\vec{r}}\,_{i}^{\alpha} = \frac{1}{\xi_0} \vec{F}_i^{\alpha}
+ \vec{\eta}_i(t) ,
\end{equation}
where the friction coefficient of an isolated particle $\xi_0 = 1.0$ and $\vec{F}_i^{\alpha}$
is the force acting on the $i_{th}$ particle of type $\alpha$,
\begin{equation}
\vec{F}_i^{\alpha}= - \nabla_i^{\alpha} \sum_{j \ne i} \sum_{\beta=1}^2
V_{\alpha \beta}\left(\left|\vec{r}\,_i^{\alpha} - \vec{r}\,_j^{\beta} \right| \right)
\end{equation}
with $\nabla_i^{\alpha}$ being the gradient operator with respect to $\vec{r}\,_i^{\alpha}$.
In Eq. (\ref{Lang}) 
the random noise $\vec{\eta}_i$ satisfies the fluctuation-dissipation 
theorem, 
\begin{equation}\label{fd}
\left\langle \vec{\eta}_i(t) \vec{\eta}_j(t') \right\rangle = 
2 D_0 \delta(t-t') \delta_{ij} \mathbf{1}.
\end{equation}
In Eq. (\ref{fd}), the diffusion coefficient $D_0 = k_B T/\xi_0$ where
$k_B$ is Boltzmann's constant and $\mathbf{1}$ is
the unit tensor.  Since the equation of motion allows for diffusive motion
of the center of mass, all the results will be presented relative to the
center of mass (\textit{i.e.} momentary positions of all the particles are always 
relative to the momentary position of the center of mass \cite{remark}).  
The results are presented in terms of the reduced units
with $\sigma_{AA}$, $\epsilon_{AA}$, $\epsilon_{AA}/k_B$, 
and $\sigma_{AA}^2 \xi_0/\epsilon_{AA}$ being the units of length, energy, 
temperature, and time, respectively. Since in these units the short-time self-diffusion 
coefficient is proportional to the temperature, in the
comparisons with the mode-coupling theory the times are rescaled to $t^* = t D_0/\sigma_{AA}^2$.

The equations of motion, Eq.~\ref{Lang}, were solved using a Heun algorithm with
a small time step of $5 \times 10^{-5}$.  To save disk space, not all the generated
configurations were saved to disk.  Thus, the short time dynamics are not
available at the lower temperatures.  We simulated the temperatures
$T = 0.44$, 0.45, 0.47, 0.50, 0.55, 0.60, 0.80, 0.90, 1.0, 1.5, 2.0, 3.0, and 5.0.
We ran equilibration runs and 4-6 production runs.  
The equilibration runs were typically twice shorter than the production runs, and the latter
were up to $6 \times 10^8$ time steps long for the lowest temperatures studied.
The results presented are averages over the production runs.

\section{\label{mct}Mode-Coupling Theory}

The mode-coupling theory leads to a
set of integro-differential equations for the coherent intermediate scattering functions 
(\textit{i.e.} dynamic partial structure factors),
\begin{equation}
S_{\alpha \beta}(q,t) = \left \langle \rho_{\vec{q}}^\alpha(t) 
\rho_{-\vec{q}}^{\beta}(0) \right \rangle 
\end{equation} 
where
\begin{equation}
\rho_{q}^{\alpha}(t) = \frac{1}{\sqrt{N_{\alpha}}} 
\sum_{i=1}^{N_{\alpha}} e^{-\mathrm{i}\vec{q}\cdot\vec{r}\,_i^{\alpha}(t)}.
\label{rho}
\end{equation}
Note that the sum in Eq. (\ref{rho}) is taken over particles of type $\alpha$.
The structure factors depend only on the magnitude of the
wave vector $|\vec{q}\,| = q$.  
The time evolution of $\rho_{\vec{q}}^\alpha(t)$ for a system of 
interacting Brownian particles is governed by the 
adjoint Smoluchowski operator \cite{Nagele}
\begin{equation}
\Omega = D_0 \sum_{\alpha=1}^2 \sum_{i=1}^{N_{\alpha}}
\left[\nabla_{i}^{\alpha} - \beta \vec{F}_i^{\alpha} 
\right]\cdot\nabla_i^{\alpha}
\end{equation}
where $\beta=1/k_B T$
and $D_0$ is the short time diffusion coefficient which is the same for both
types of particles. We set $D_0 = 1.0$ for these calculations.

The mode-coupling equations governing the time evolution of 
the coherent intermediate scattering functions for Brownian mixtures 
have been derived by N\"agele \textit{et al.} \cite{Nagele1}:
\begin{eqnarray}
\frac{\partial}{\partial t} \mathbf{S}(q,t) & = & - q^2 D_0\:
\mathbf{S}^{-1}(q)\: \mathbf{S}(q,t)\nonumber\\
& & - \int_{0}^{t} \mbox{d}u\ \mathbf{M}(q,t-u) \frac{\partial}{\partial u} 
\mathbf{S}(q,u)\nonumber\\
\label{coheq}
\end{eqnarray} 
where
\begin{equation}
\mathbf{S}(q,t) = \begin{pmatrix} S_{AA}(q,t) & S_{AB}(q,t) \\ S_{BA}(q,t) & S_{BB}(q,t) 
\end{pmatrix},
\end{equation}
is the matrix of coherent intermediate scattering functions,
and $\mathbf{M}$ is the matrix of memory functions,
\begin{eqnarray}
M_{\alpha \beta}(\vec{q},t) & = & \frac{V D_0}{32 \pi^2 \sqrt{N_\alpha N_\beta}} \nonumber \\
& & \times \sum\limits_{l,l^\prime,m,m^\prime}
\int \mathrm{d}\vec{k}V_{\alpha l m}(\vec{q},\vec{k})
V_{\beta l^\prime m^\prime}(\vec{q},\vec{k}) \nonumber \\
& & \times\ S_{m m^\prime}(|\vec{q} - \vec{k}|,t) S_{l l^\prime}(k,t)
\label{cohmem}
\end{eqnarray}
where the vortex
\begin{eqnarray}
V_{\alpha l m}(\vec{q},\vec{k}) & = & \frac{\vec{q} \cdot \vec{k}}{q} 
\delta_{\alpha m}C_{\alpha l}(k)\nonumber\\ 
& & +
\frac{\vec{q} \cdot (\vec{q} - \vec{k})}{q} \delta_{\alpha l}C_{\alpha m}(| \vec{q} - \vec{k} |).
\label{vortex}
\end{eqnarray}
In Eq. (\ref{vortex}) the matrix $\mathbf{C}(q)$ is defined through the
Ornstein-Zernike matrix equation 
\begin{equation}
\mathbf{S}^{-1}(q) = \mathbf{1} - \mathbf{C}(q),
\label{direct}
\end{equation}
where $\mathbf{1}$ is the unit tensor.
The mode-coupling theory equations allows one to calculate the time evolution of 
$S_{\alpha \beta}(q,t)$ with only the time independent quantity
$S_{\alpha \beta}(q) = S_{\alpha \beta}(q,0)$ 
as an input. 

The incoherent intermediate scattering functions 
\begin{equation}
F_{\alpha}^s(q,t) = 
\left\langle e^{\mathrm{i} \vec{q}\cdot 
\left[ \vec{r}\,_{i}^{\alpha}(t) - \vec{r}\,_{i}^{\alpha}(0) \right]} \right \rangle
\label{self}
\end{equation}
are calculated using as input the coherent intermediate scattering function and
the partial structure factors.
N\"agele \textit{et al.} \cite{Nagele1} derived the equations governing the time evolution of 
$F_{\alpha}^s(q,t)$ for Brownian mixtures, 
\begin{eqnarray}
\frac{\partial}{\partial t}F_{\alpha}^s(q,t) & = & -q^2 D_0 F_{\alpha}^s(q,t) \nonumber\\
& & - \int_{0}^{t} \mbox{d}u M_{\alpha}^s(q,t-u) \frac{\partial}{\partial u} F_{\alpha}^s(q,u)
\label{selfeq} 
\end{eqnarray}
where the memory function
\begin{eqnarray}
M_{\alpha}^s(q,t) & = & \frac{D_0 V}{\left(2 \pi \right)^2 N_{\alpha}} 
\int \mbox{d}\vec{k} \left(\frac{\vec{q}\cdot\vec{k}}{q}\right)^2 
F_{\alpha}^s(|\vec{q} - \vec{k} |) \nonumber\\
& & \times \sum_{\delta \delta^{\prime}}C_{\alpha \delta}(k) 
S_{\delta \delta^{\prime}}(k,t)C_{\alpha \delta^{\prime}}(k)
\label{selfmem}
\end{eqnarray}

For short times the integrals involving the memory function 
in Eqs. (\ref{coheq}) and (\ref{selfeq}) are 
approximately zero, therefore
\begin{equation}
\mathbf{S}(q,t) \approx \exp\left[-q^2 D_0\,\mathbf{S}^{-1}(q)\,t\right]\,\mathbf{S}(q),
\label{cohapprox}
\end{equation}
\begin{equation}
F_{\alpha}^s(q,t) \approx \exp\left[-q^2 D_0 t\right].
\label{selfapprox}
\end{equation}
In Eq. (\ref{selfapprox}) we used $F_{\alpha}^s(q,0) = 1.0$.

According to Eqs. (\ref{cohapprox}-\ref{selfapprox}), 
at short times the particles undergo diffusive motion 
with a diffusion coefficient $D_0$.  The effect of the memory function
is to provide a feedback mechanism which produces a ``caging'' of the particles.  
For temperatures below the
transition temperature $T_c$, there is structural arrest and
the particles do not escape
their cage.  This results in a non-zero value of 
$\mathbf{S}(q,t)$ and $F_{\alpha}^s(q,t)$ as $t \rightarrow \infty$.
For temperatures close to but above $T_c$, there is a plateau region in the
log-log plot of the mean square displacement and the log-linear plot of the 
intermediate scattering functions. At long times the motion of the particles
is again diffusive with a temperature dependent diffusion coefficient
$D \ll D_0$.  

We calculate the input to the mode-coupling equations, \textit{i.e.}~the 
partial static structure factors $S_{\alpha \beta}(q)$,
directly from the Brownian dynamics
simulations. Because of the finite size of the simulation box, the 
magnitude of the smallest wave vector calculated is $2 \pi/L$,
where $L$ is the length of the simulation box.
We extrapolated $S_{\alpha \beta}(q)$ 
to zero by fitting the first few wave vectors
to a polynomial.  The method used to calculate the integrals
of the memory 
functions require that the partial structure factors are 
known at equally spaced wave vectors (see the appendix for
details on the numerical procedure implemented in this work).  
The structure factor for these wave vectors
were determined by fitting the partial structure
factors determined from the simulations to a cubic spline. 

To solve the mode-coupling equations, we used 
300 equally spaced wave vectors from $q=0$ to $q=40$
with the first wave vector $q_0 = 0.2/3$.  
We performed a few calculations with
larger cutoffs for the integral and/or a finer grid of
wave vectors.  The difference in the values of the 
calculated intermediate scattering functions 
was at most 5\% and less in most cases.
This difference results in a less than 5\% difference in
the self-diffusion coefficient and the 
incoherent intermediate scattering function's relaxation
time.

The partial structure factors for temperatures which were
not directly simulated were calculated by a quadratic 
polynomial interpolation between the points at 
three adjacent temperatures.  For most interpolated
temperatures, it is possible to use two different sets
of temperatures to determine the structure factor
at the interpolated temperature.  Using different temperature
ranges changed the value of $\mathbf{S}(q,t)$ by as much as
1\%, but by less in most cases.  
At a few temperatures we also used a 
linear interpolation between two adjacent
temperatures or a cubic interpolation using the 
four closest temperatures.  The difference in the
results were less than 2\% using these different interpolation 
schemes. We conclude that the results depends little
on the interpolation scheme.

First, we solved the mode-coupling equations for the
coherent intermediate scattering functions until all the 
scattering functions decayed to zero or to a non-zero constant.  Then the 
coherent scattering functions and the structure factor were used as the inputs for
the calculation of the incoherent scattering functions, which were solved for
the same times as for the coherent intermediate scattering functions.  

Since the mode-coupling equations have to be solved for many 
decades in time, specialized techniques have been developed.
We describe the method used to solve the mode-coupling equations
in the appendix.  The algorithm was first described  
by Fuchs \textit{et al.} \cite{FuchsMCT}, and then, in more detail, by Miyazaki, 
Bagchi and Yethiraj \cite{Miyazaki}. 
It is an iterative technique which calculates
the coherent and the incoherent scattering functions
from $t = i \Delta t$ where
$i \in \{N+1,\dots,2N\}$, assuming that the scattering functions are known 
for $t = j \Delta t$ where $j \in \{1,\dots,N\}$.  Then the
time step $\Delta t^{\prime} = 2 \Delta t$ is doubled, and the values of the 
scattering functions for $t = i \Delta t^{\prime}$ where $i \in \{1, \dots,N\}$
are determined from the values of the scattering functions for
$t = j \Delta t$ where $j \in \{1, \dots, 2N\}$. We begin
the calculation for an initial time step of $10^{-8}$. For the first
set of times, the scattering functions are not known.
We used the approximation given by Eq. (\ref{cohapprox}) and (\ref{selfapprox})
to supply the values of the coherent and the incoherent scattering
functions for the initial $N$ times.  To check this approximation, we also used a more time 
consuming procedure to solve the
mode-coupling equations for the initial $N$ times.  In this procedure, 
the integrals of the memory function
are included for $t > 10^{-8}$.   
The difference in all calculated quantities was less than 0.5\%.     

\section{\label{transT}Transition Temperature: Mode-Coupling Theory \textit{vs.} Simulations}

\begin{figure}
	\includegraphics[scale=0.25]{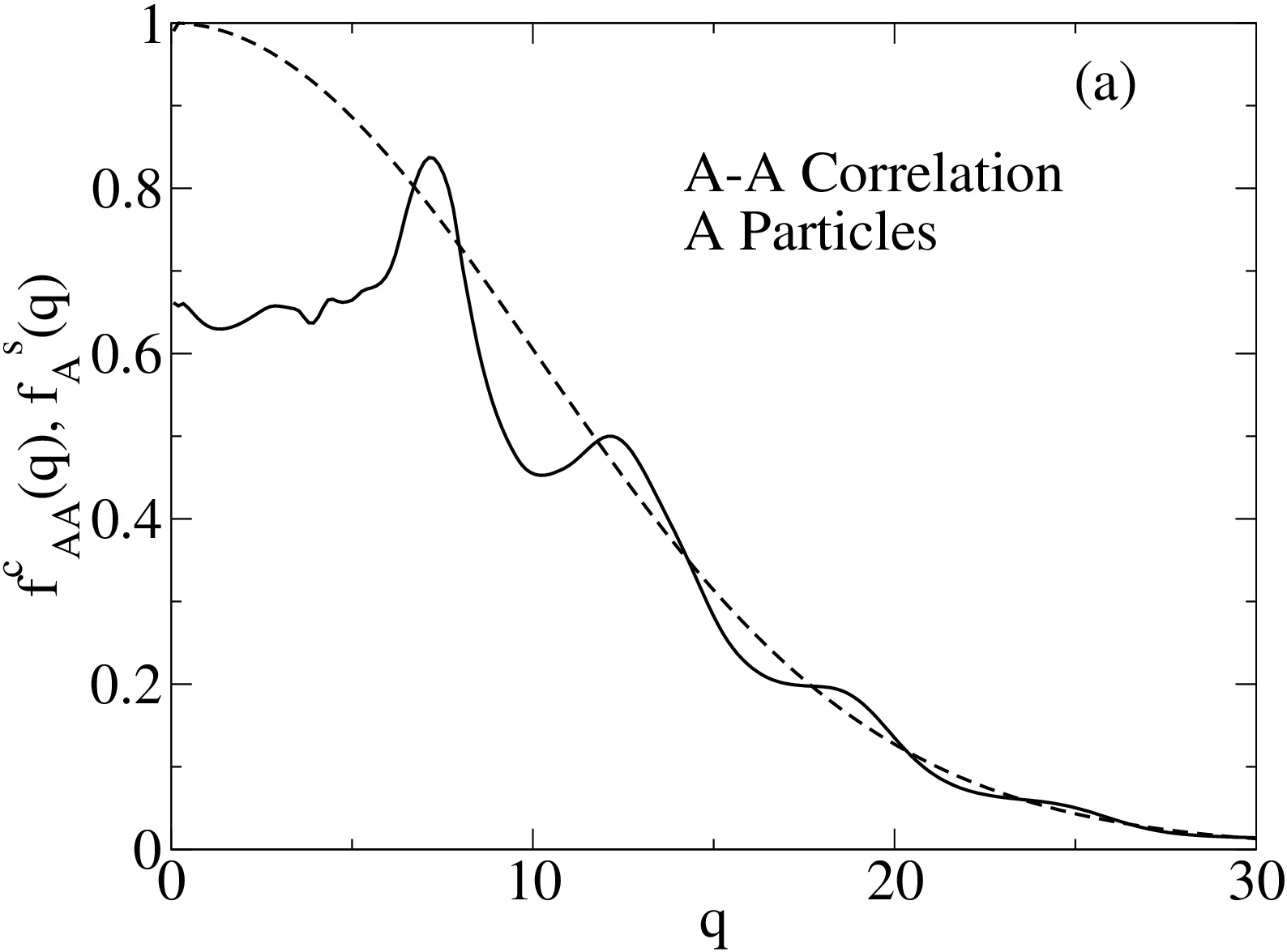}\\[0.25cm]
	\includegraphics[scale=0.25]{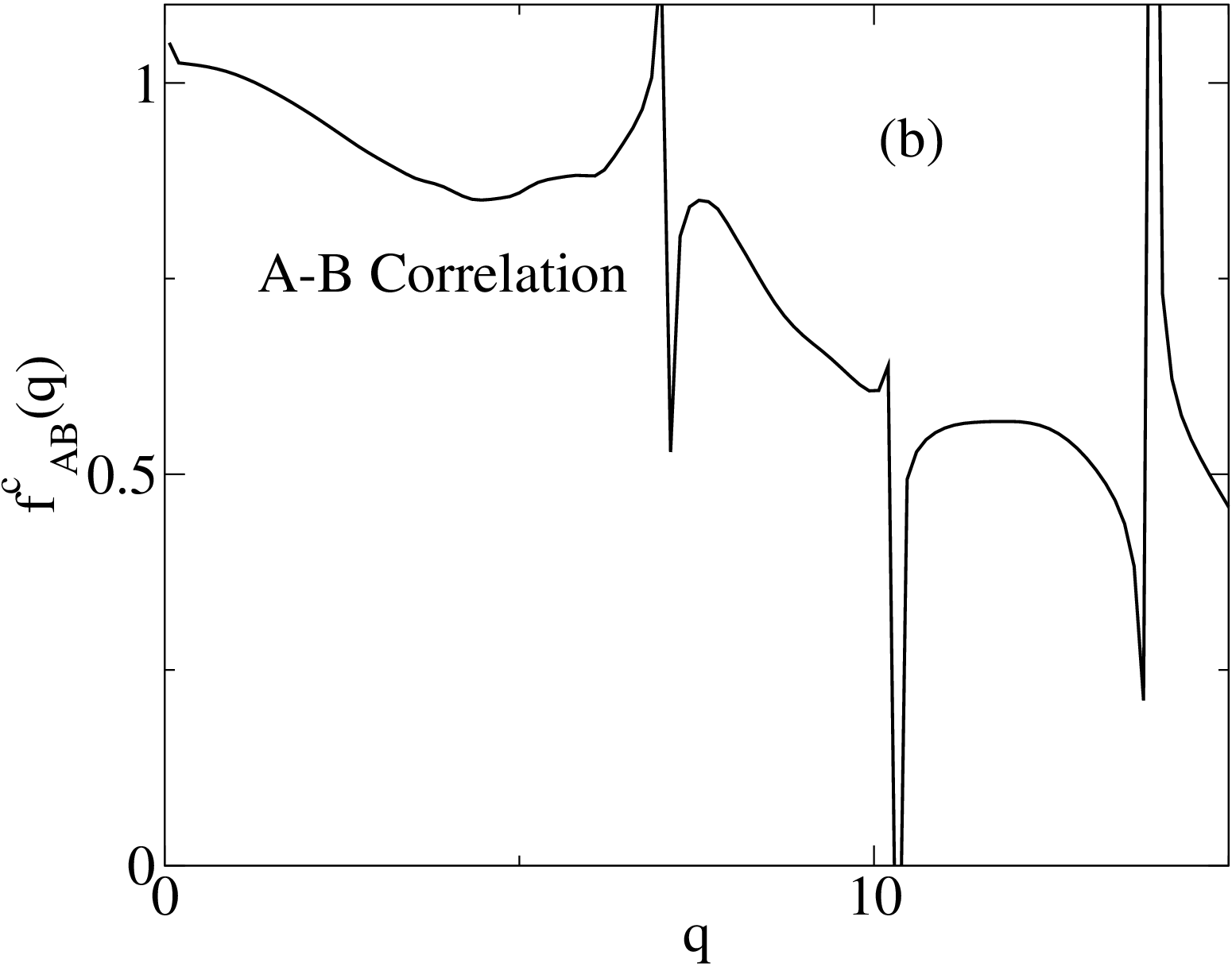}\\[0.25cm]
	\includegraphics[scale=0.25]{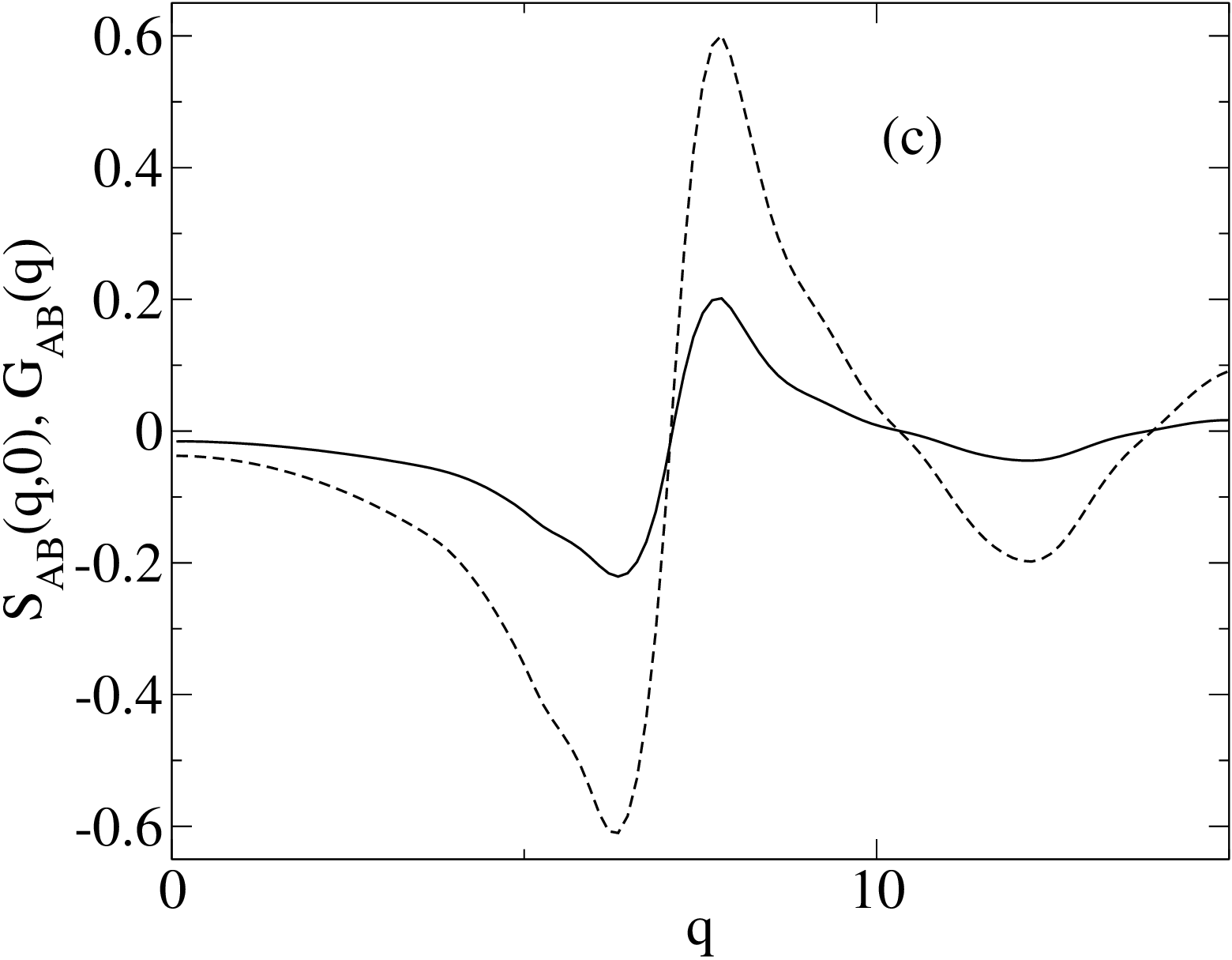}\\[0.25cm]
	\includegraphics[scale=0.25]{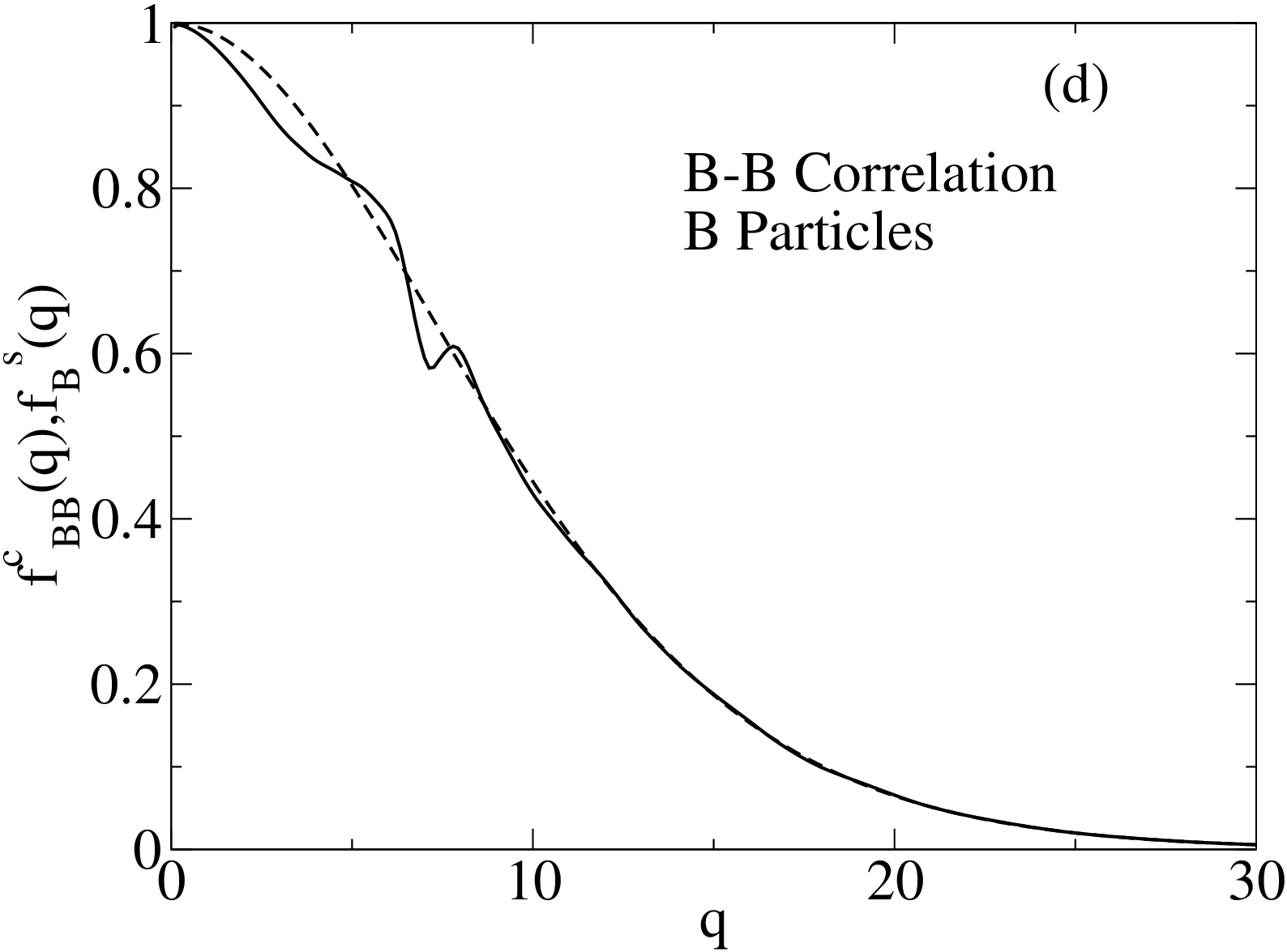}
\caption{\label{neplot}(a) Non-ergodicity parameter at $T_c$ for the A particles.
Solid line: coherent correlator, A-A particles. Dashed line: incoherent correlator,
A particles. (b) Non-ergodicity parameter at $T_c$: coherent correlator, A-B particles.
(c) Dashed line: A-B partial structure factor, $S_{AB}(q)$. Solid line:
$G_{AB}(q) = S_{AB}(q;t\rightarrow\infty)$ at $T_c$. 
(d) Non-ergodicity parameter at $T_c$ for the B particles.
Solid line: coherent correlator, B-B particles. Dashed line: incoherent correlator,
B particles. }
\end{figure}

The mode-coupling theory predicts an ergodicity breaking transition when 
$S_{\alpha \beta}(q,t \rightarrow \infty) \ne 0$ at some critical 
temperature $T_c$.  
To find the transition temperature we calculate
$\mathbf{G}(q) = \lim_{t \rightarrow \infty} \mathbf{S}(q,t)$
as a function of temperature.
We followed the 
procedure used in previous calculations of the
non-ergodicity parameter \cite{BarratLatz,Nauroth,Barrat} to
calculate $\mathbf{G}(q)$.
First, we took the
Laplace transform of Eq.~\ref{coheq} to get
\begin{eqnarray}
\lefteqn{\left[ z + z\mathbf{M}(q,z) + q^2 D_0 \mathbf{S}^{-1}(q) \right]\mathbf{S}(q,z) =} 
\hspace{5cm} \nonumber \\
 & & \mathbf{S}(q) + \mathbf{M}(q,z)\mathbf{S}(q) \nonumber \\
\end{eqnarray}
where $f(z)$ denotes the Laplace transform of $f(t)$, $f(z) = \int_0^{\infty} e^{-zt}f(t)$.  
To find the long time limit of $\mathbf{S}(q,t)$ we utilized the relationship 
$\lim_{t \rightarrow \infty} f(t) = \lim_{z \rightarrow 0} z f(z)$ and derived 
an equation for $\mathbf{G}(q)$. This equation was then solved using the
following iterative procedure,
\begin{equation}
\mathbf{G}^{(i+1)}(q) = \left[ q^2 D_0 \mathbf{1} + \mathbf{S}(q)\mathbf{M}^{(i)}(q) \right]^{-1} 
\mathbf{S}(q)\mathbf{M}^{(i)}(q)\mathbf{S}(q),
\label{nepeq} 
\end{equation}
where $\mathbf{M}^i(q)$ is calculated from $\mathbf{G}^i(q)$.
The non-ergodicity parameters are defined as
$f_{\alpha \beta}^c(q) = G_{\alpha \beta}(q)/S_{\alpha \beta}(q)$.
Using the same method, we found the non-ergodicity parameters 
for the incoherent intermediate scattering functions, 
$F_{\alpha}^s(q,t \rightarrow \infty) = f_{\alpha}^s(q)$.
The iterative procedure 
for the non-ergodicity parameter for the 
self correlation functions is
\begin{equation}
\frac{f_{\alpha}^{s(i+1)}}{1 - f_{\alpha}^{s(i+1)}} = \frac{M_{\alpha}^{s(i)}}{q^2 D_0}, 
\label{neseq} 
\end{equation}
where $M_{\alpha}^{s(i)}$ depends on $f_{\alpha}^{s(i)}$ 
and the solution to Eq. (\ref{nepeq}), $G_{\alpha \beta}$.

The input to Eq. (\ref{nepeq}) are the partial
structure factors for a temperature $T$.  The iterative procedure was
followed until it was found that either $\mathbf{G}(q)$ is zero for 
all $q$ or $\mathbf{G}(q)$ is nonzero and does not change anymore for all $q$.  
If $\mathbf{G}(q)$ is zero, then $T > T_c$ 
otherwise $T \le T_c$.
By trying different structure factors for different temperatures
the transition temperature can be determined to arbitrary precision.  
This method is preferred to finding the
transition temperature by calculating the full time dependence of $\mathbf{S}(q,t)$,
since calculating $\mathbf{G}(q)$ is around 20 times faster and reduces
the calculation of the non-ergodicity parameter to 
hours instead of days.  Once $T_c$ was 
found, we calculated the non-ergodicity parameter by solving the 
mode-coupling equations for $\mathbf{S}(q,t)$ and $F_{\alpha}^{s}(q,t)$ 
at $T_c$.  The non-ergodicity
parameters found using the two methods agree.

The non-ergodicity parameters at $T_c$ 
are plotted in figure \ref{neplot} for the
AA, AB, and BB correlators 
(solid lines), and for the incoherent
intermediate scattering functions for the A and B particles (dashed lines).
The results are similar to what was obtained by Nauroth and Kob \cite{Nauroth}.
All the non-ergodicity parameters are nonzero for $q=0$, but approach
zero for large $q$.  The small features for $q < 5.0$ for the 
AA non-ergodicity parameters
are numerical and do not represent additional features in
the non-ergodicity parameter. 
The division of $G_{AB}(q)$ by $S_{AB}(q)$ for the AB non-ergodicity parameter causes numerical
problems when $S_{AB}(q) \approx 0$.  This is seen as large
spikes in $f^c_{AB}$.  In the insert we show 
the input to Eq. (\ref{nepeq}), $S_{AB}(q)$, (dashed line) and
the results of the calculation $G_{AB}(q)$ (solid line) at $T_c$. An extensive
comparison of the mode-coupling theory predictions to the simulation results
for the non-ergodicity parameters 
has already been conducted \cite{Nauroth}, and we do not repeat it here.    

We determined a transition temperature $T_c^{theory} = 0.9515$,
which is 3\% higher than the ergodicity breaking
temperature $T_c^{theory, NK} = 0.922$ determined by
Nauroth and Kob \cite{Nauroth}.  Foffi \textit{et al.}~\cite{Foffi}
showed that 
small differences in the structure factor
can result in large differences in the critical packing fraction for
a binary system of hard spheres.
They found that the critical packing fraction was around 
5\% higher if the partial structure factors were determined from 
the results of Newtonian dynamics simulations instead of
using the Percus-Yevick approximation, 
even though there was little difference in the 
partial structures factors. 

The transition temperature predicted by the theory is around a factor of two larger 
than the commonly accepted mode-coupling temperature 
inferred from simulations \cite{KobAndersen,Nauroth} 
of the same system, $T_c^{sim} = 0.435$. It should be emphasized that, 
in contrast to the transition predicted by the mode-coupling theory, only a crossover 
in the dynamics is observed 
in simulations \cite{KobAndersen,SzamelFlenner,FlennerSzamel2}. 
Furthermore, recently we argued 
that there is some arbitrariness regarding the mode-coupling temperature inferred from simulations 
\cite{FlennerSzamel2}. The mode-coupling temperature is usually obtained by fitting the 
simulation results for the characteristic decay time of the intermediate scattering function
and the diffusion
coefficient to a power law $a(T-T_c)^{\gamma}$.  The 
transition temperature obtained in this manner depends on the temperature
range used in the fit. As argued in Ref. \cite{FlennerSzamel2}, around  
the commonly accepted value of the mode-coupling temperature, $T_c^{sim} = 0.435$, 
the relaxation mechanism changes from high temperature diffusive motion to 
low temperature hopping-like motion.

\section{\label{scatt} Incoherent Intermediate Scattering Functions}

\begin{table}
\caption{\label{redtemp}The reduced temperatures and their corresponding
temperature in the Brownian dynamics simulations and the mode-coupling
theory calculations.}
\begin{ruledtabular}
\begin{tabular}{ccc}
$\epsilon$ & BD Temperature & MCT Temperature \\
\hline 
0.0000 & 0.435& 0.9515 \\
0.0115 & 0.44 & 0.9624 \\
0.0345 & 0.45 & 0.9843 \\
0.0805 & 0.47 & 1.0281 \\
0.1495 & 0.50 & 1.0937 \\
0.2644 & 0.55 & 1.2030 \\
0.3793 & 0.60 & 1.3124 \\
0.8391 & 0.80 & 1.7499 \\
1.0690 & 0.90 & 1.9686 \\
1.2989 & 1.00 & 2.1873 \\
2.4483 & 1.50 & 3.2810 \\
3.5977 & 2.00 & 4.3747 \\
5.8966 & 3.00 & 6.5621\footnotemark[1] \\
10.494 & 5.00 & 10.937\footnotemark[1] \\
\end{tabular}
\end{ruledtabular}
\footnotetext[1]{mode-coupling equations were not solved for these temperatures.}
\end{table}  

We calculate the time dependence of the coherent
and the incoherent intermediate scattering functions from the
mode-coupling equations with the 
partial structure factors determined from the simulations as input. 
In this section we compare predictions of the mode-coupling theory 
with results of Brownian dynamics simulations at the
same reduced temperature $\epsilon = (T - T_c)/T_c$. In table \ref{redtemp} we
list the reduced temperatures $\epsilon$, the
corresponding temperatures for the Brownian dynamics simulations,
and the temperatures which were used in the mode-coupling theory
calculations.

\begin{figure}
	\includegraphics[scale=0.25]{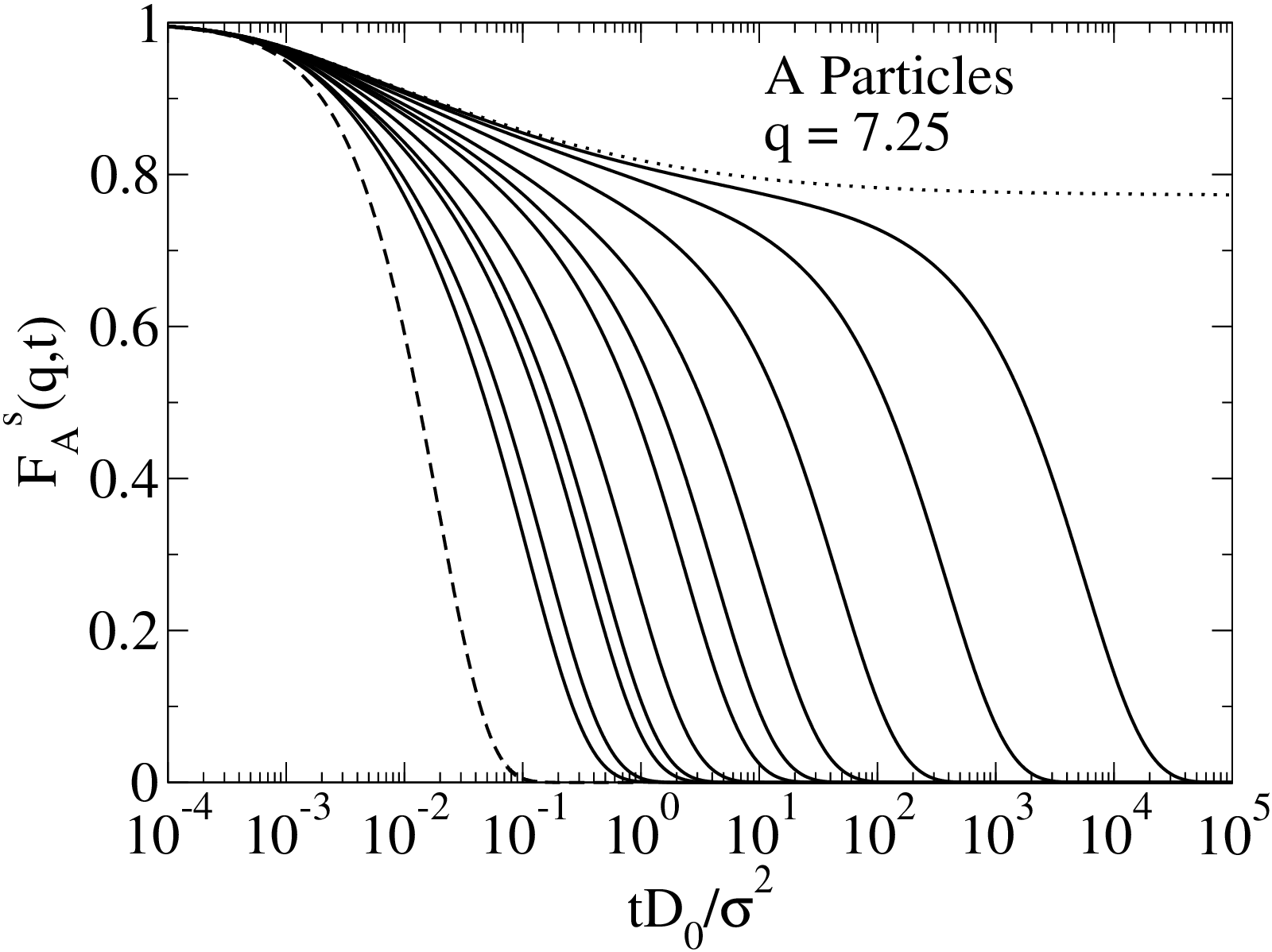}\\[0.25cm]
	\includegraphics[scale=0.25]{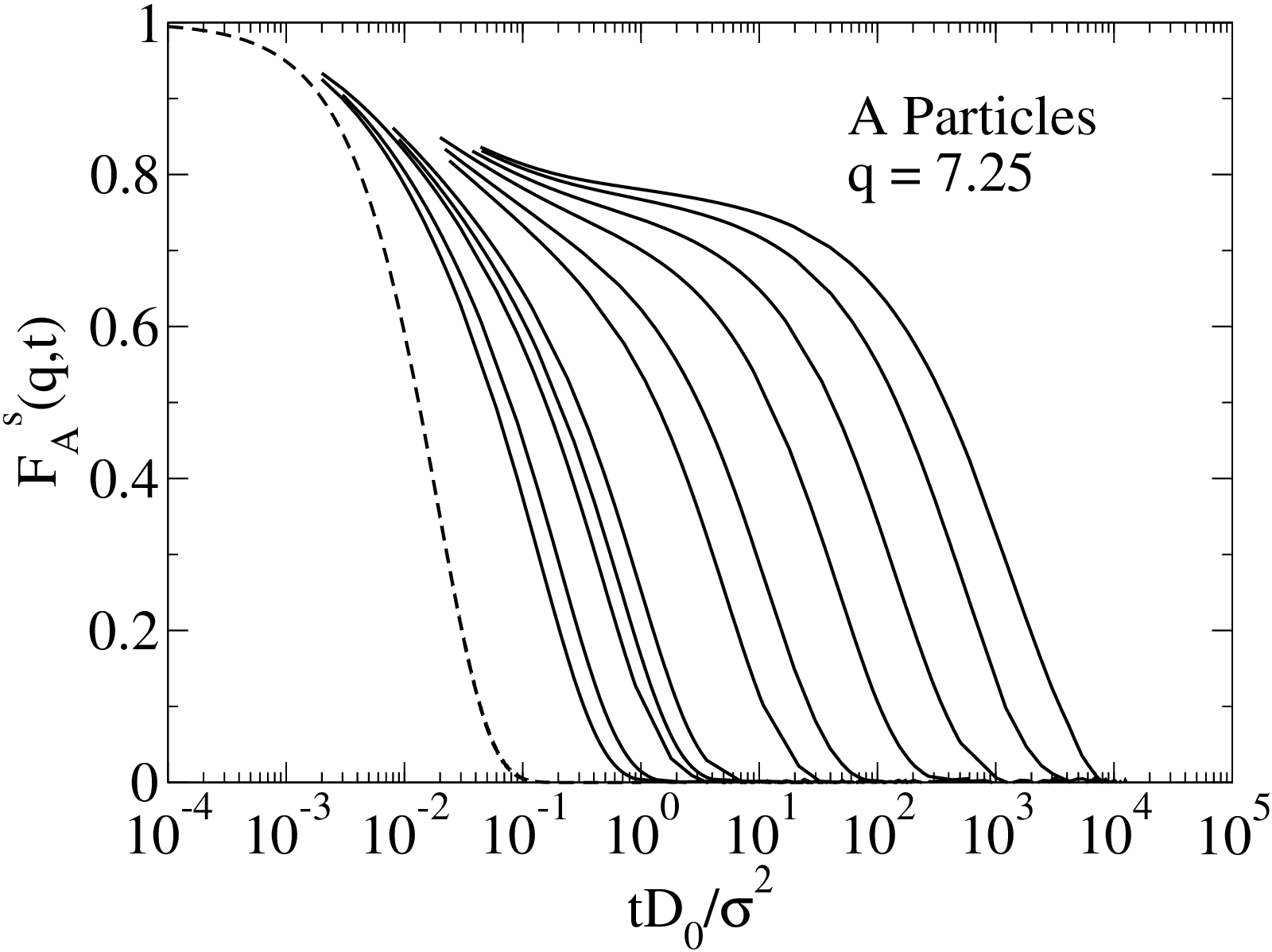}
\caption{\label{scattfig}Incoherent intermediate scattering function for 
the A particles for $q=7.25$. (a) Predicted by the mode-coupling theory.
(b) Calculated from the Brownian dynamics simulations.  
The scattering functions are shown for the
same reduced temperatures $\epsilon = (T-T_c)/T_c$.  The reduced temperatures 
are 3.5977, 2.4483, 1.2989, 1.0690, 0.8391, 0.3793, 0.2644, 0.1494, 0.0805, 0.0345, 0.0115
listed from left to right. The dashed line in 
both figures correspond to the limit of non-interacting particles.  The
dotted line in figure (a) is the incoherent intermediate scattering function
calculated at $T_c^{theory}$.}
\end{figure}

Incoherent intermediate scattering functions for the A particles are shown in
Fig.~\ref{scattfig}a and \ref{scattfig}b 
calculated using the mode-coupling theory and 
obtained from the Brownian dynamics simulations, respectively.
Both figures show the incoherent intermediate scattering function at $q = 7.25$, 
which is around the first peak of $S_{AA}(q)$.  
The scattering function for the mode-coupling theory was
calculated through a linear interpolation of nearby
scattering functions which were calculated directly from
the mode-coupling equations.  The dotted line in
Fig.~\ref{scattfig}a is the mode-coupling results at
$T_c^{theory} = 0.9515$. The dashed line in each figure shows the incoherent
scattering function for non-interacting particles, Eq.~(\ref{selfapprox}).
The mode-coupling theory correctly predicts the
two step decay of the intermediate scattering functions,
and the plateau observed in the log-linear plot of the
scattering functions. 

\begin{figure}
	\includegraphics[scale=0.25]{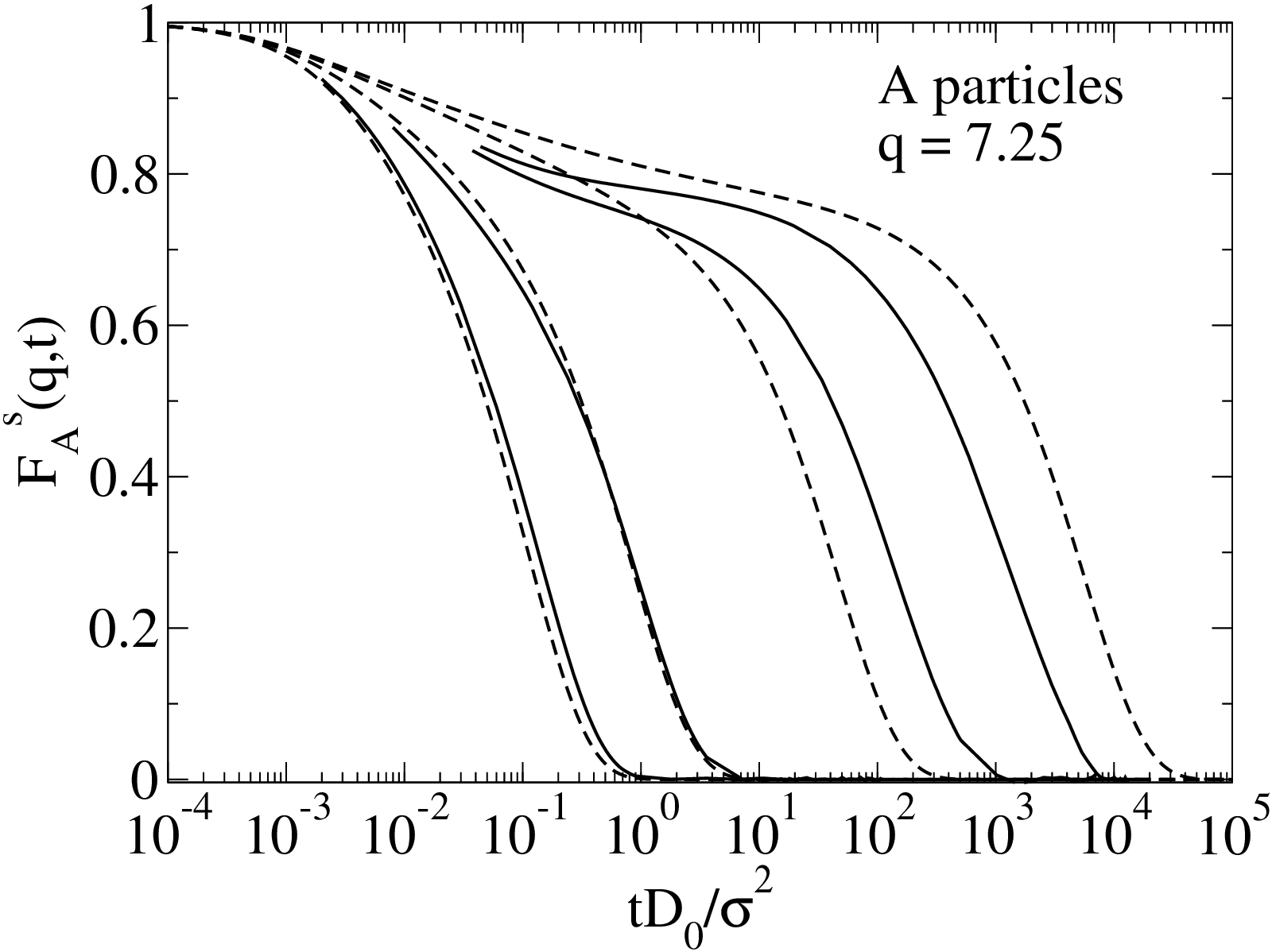}\\[0.25cm]
	\includegraphics[scale=0.25]{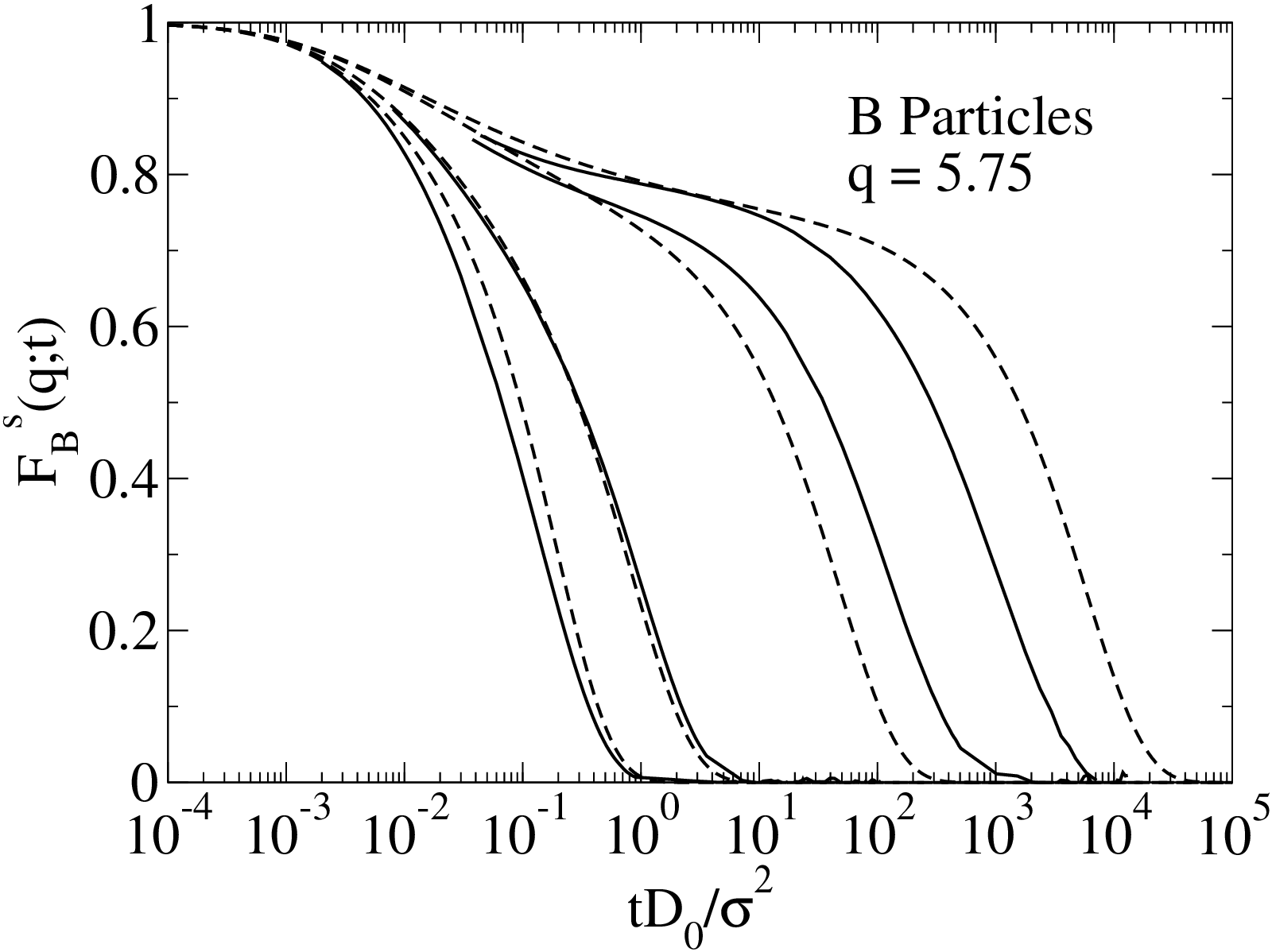}
\caption{\label{scattcomp}The incoherent intermediate scattering
function for the A and B particles predicted by the
mode-coupling theory (dashed lines) and calculated
from the Brownian dynamics simulations (solid lines)
for $\epsilon = 3.598$, 0.839, 0.0805 and 0.0115 listed from
left to right.}
\end{figure}

In Fig.~\ref{scattcomp}a we 
compare the incoherent intermediate scattering functions 
for $\epsilon = 3.5977$, 0.8391 0.0805, and 0.0115 for the A particles
at $q = 7.25$, and in Fig.~\ref{scattcomp}b we show the comparison for the
same reduced temperatures for the B particles at $q = 5.75$.  At the
higher reduced temperatures there is very good agreement between
the mode-coupling calculations and the Brownian dynamics simulations.
For $0.0345 < \epsilon < 0.8391$ the characteristic decay time 
of the scattering functions calculated from the mode-coupling theory
is less than that of the Brownian dynamics
simulation.  For reduced temperatures equal to and below 0.0345, the
mode-coupling theory predicts a longer decay time for the self
intermediate scattering functions for this value of $q$.  
However, the shape of the 
incoherent intermediate scattering functions are similar in the
$\alpha$ relaxation region.

Kob, Nauroth, and Sciortino \cite{KobNauroth} compared
the self intermediate scattering functions obtained from
Newtonian dynamics simulations and predicted by the mode-coupling theory. 
The input temperature in the mode-coupling theory was adjusted so that
the relaxation time of $S_{AA}(q,t)/S_{AA}(q)$ was correctly reproduced
for one wave vector around the first peak of $S_{AA}(q)$.
The procedure followed by Kob \textit{et al.}\ requires that
the characteristic decay time
is the same in the simulations and the mode-coupling theory calculations.
Kob \textit{et al.}\ observed that the 
shape of the scattering functions and its wave vector dependence
is accurately described by the theory for reduced temperatures
above 0.071.  We discuss the wave vector
dependence of the relaxation time in section \ref{vanhove}.  

\begin{figure}
	\includegraphics[scale=0.25]{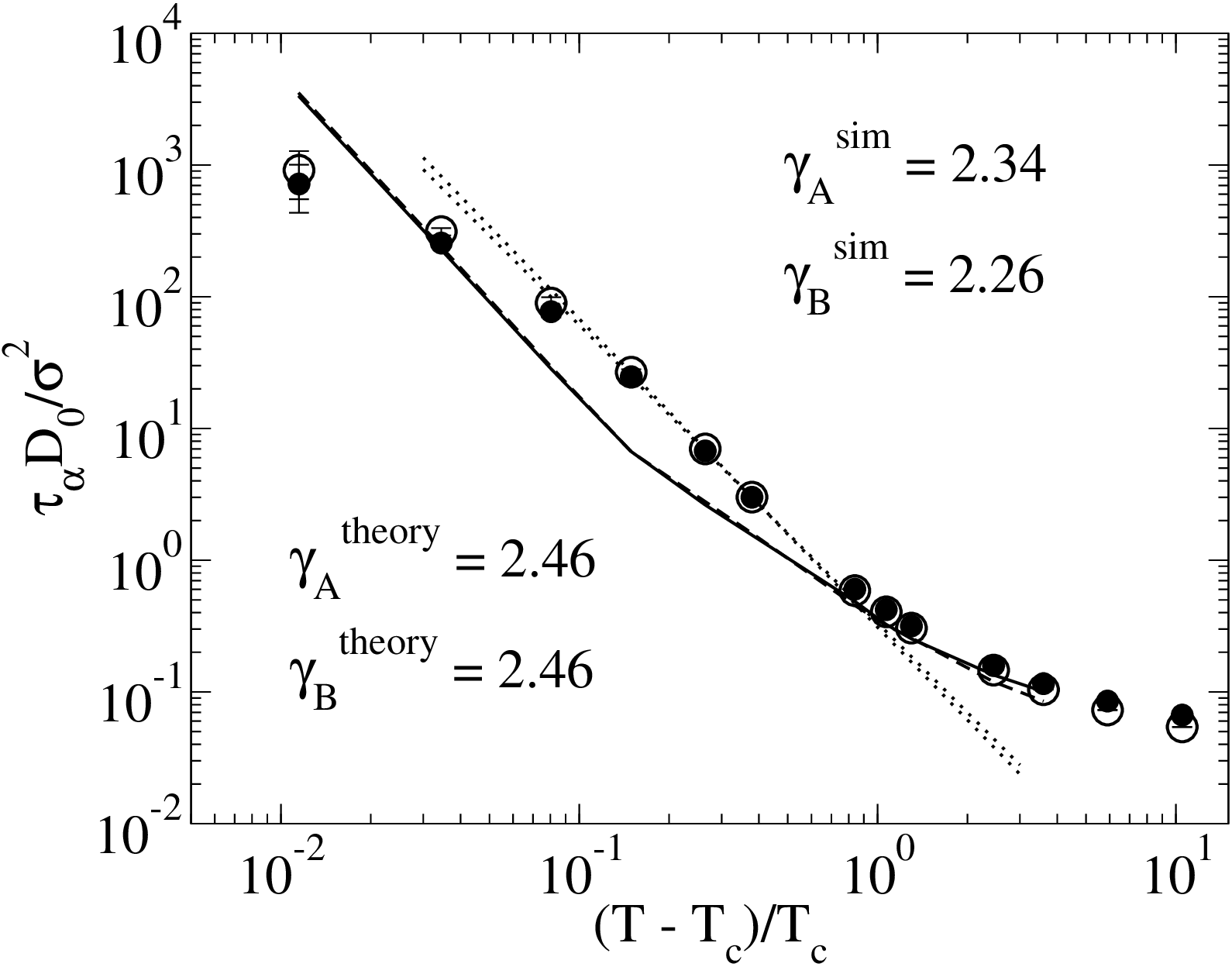}
\caption{\label{alpha}The $\alpha$ relaxation time 
calculated from the Brownian dynamics simulations (symbols) and
predicted by the mode-coupling theory (solid and dashed lines) for
the the A and B particles.  
The A particles are represented by the closed symbols and the solid line.
The B particles are represented by the open symbols and the dashed line. 
The lines nearly overlap.  The dotted lines are 
fits of the simulation data to the function $a[(T-T_c)/T_c]^{-\gamma}$.
The exponents to the power law fits are given in the figure along with
the exponents to power law fits to the predictions of the mode-coupling
theory.}
\end{figure}

The mode-coupling theory predicts power law divergence of the
characteristic decay time of the intermediate scattering functions.  Specifically, we define
the $\alpha$ relaxation time as the time when the incoherent intermediate scattering 
function decays to
$e^{-1}$ of its initial value, 
$F_{\alpha}^s(q,\tau_{\alpha}) = e^{-1}$.  In Fig.~(\ref{alpha}) we show  
the $\alpha$ relaxation time for the A and B particles
as a function of reduced temperature.
We fit the $\alpha$ relaxation time to the function $a[(T-T_c)/T_c]^{-\gamma}$.  
We fit the Brownian dynamics results
to reduced temperatures from 0.8391 to 0.1495, 
which corresponds to the same reduced temperatures fit in 
an earlier work \cite{FlennerSzamel2}.  We used this range
of temperatures since this is the temperature range 
in which a power law fits the $\alpha$ relaxation time well
for a transition temperature of 0.435.
For the mode-coupling theory calculations, we fit the power law 
to reduced temperatures equal to and below 0.0345.  
The exponents in the power law fits are given
in the figure.  

The exponents from the mode-coupling calculations are
close to but slightly larger than the exponents found from the simulations.  
Note that the exponent obtained from solving 
the mode-coupling equations for the 
single component hard
sphere system \cite{Fuchs} is the same as the exponents predicted by the mode 
coupling coupling theory for the binary Lennard-Jones system. 
The exponents determined
from simulations are slightly different
than the exponents reported in an earlier work \cite{FlennerSzamel2}, 
since we are fitting $\tau_{\alpha}D_0$ here and we fit $\tau_{\alpha}$ in the other work.
We find, as other authors have also observed \cite{KobAndersen}, that the 
exponents determined from the simulations are different for the A and the B particles.  
However, this difference is within the 3\% uncertainty in the exponents.
Note that the mode coupling theory predicts the same exponents for the A and B
particles, and the same exponents for the $\alpha$ relaxation time and the 
self diffusion coefficient.

Furthermore, the predictions of the mode-coupling theory 
follow the asymptotic power law behavior for
reduced temperatures smaller than $\epsilon \approx 0.08$. 
This is in contrast with the results of simulations, which 
can only be fit to power laws in a restricted range $0.1495 < \epsilon < 0.8391$. 
On the other hand, it has been
observed in experiments that the power law behavior of
$\tau_{\alpha}$ is valid for reduced temperatures
up to 0.5~\cite{Du}. 

\section{\label{msd} Mean Square Displacement}

To derive the mode-coupling theory predictions for the mean square displacement 
we use the small $q$ expansion of the incoherent intermediate  
scattering function \cite{Kaur},
\begin{equation}
F_{\alpha}^s(q,t) = \sum_{n=0}^{\infty} (-1)^n \frac{q^{2n}}{(2n+1)!} 
\left< \delta r_{\alpha}^{2n}(t) \right>,
\label{expand}
\end{equation}
where $\left< \delta r_{\alpha}^{2n}(t) \right> = 
\left< |\vec{r}\,_i^{\alpha}(t) - \vec{r}\,_i^{\alpha}(0) |^{2n} \right>$.
Inserting Eq.~(\ref{expand}) into the equation for the incoherent intermediate 
scattering function, Eq.~(\ref{selfeq}), and expanding $M_{\alpha}^s(q,t)$ in a
Taylor series
\begin{equation}
M_{\alpha}^s(q,t) = \mathcal{M}_{\alpha}^0(t) + q^2 \mathcal{M}_{\alpha}^2(t) + 
q^4 \mathcal{M}_{\alpha}^4(t) + 
\dots
\end{equation}
results in the equation of motion 
\begin{equation}
\frac{\partial}{\partial t} \left< \delta r_{\alpha}^2(t) \right> 
= 6 D_0 - \int_{0}^{t} \mbox{d}u \mathcal{M}_{\alpha}^0(t-u) \frac{\partial}{\partial u} 
\left< \delta r_{\alpha}^2(u) \right>
\label{msdeq}
\end{equation}
where 
\begin{eqnarray}
\mathcal{M}_{\alpha}^0(t) & = & \lim_{q \rightarrow 0} M_{\alpha}^s(q,t) \nonumber \\
& = & \frac{D_0 V}{6 \pi^2 N_{\alpha}} \int_0^{\infty} \mbox{d}k\: k^4\: F_{\alpha}^s(k,t) 
\nonumber \\
& & \times  
\sum_{\delta \delta^{\prime}}C_{\alpha \delta}(k) S_{\delta \delta^{\prime}}(k,t)
C_{\alpha \delta^{\prime}}(k).
\label{mzero}
\end{eqnarray}
This equation was solved using a variation of the procedure described
in the appendix.  

\begin{figure}
	\includegraphics[scale=0.25]{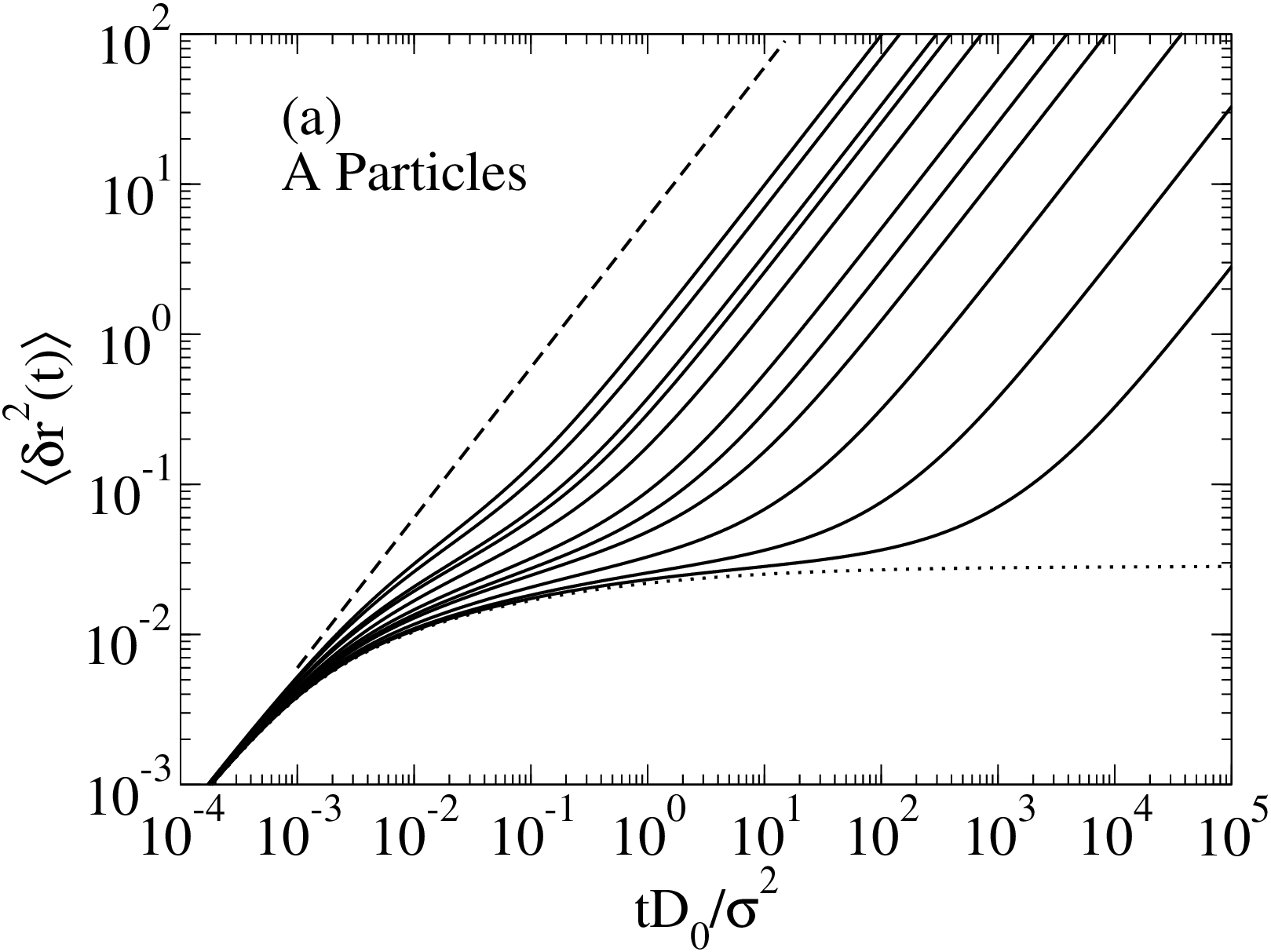}\\[0.25cm]
	\includegraphics[scale=0.25]{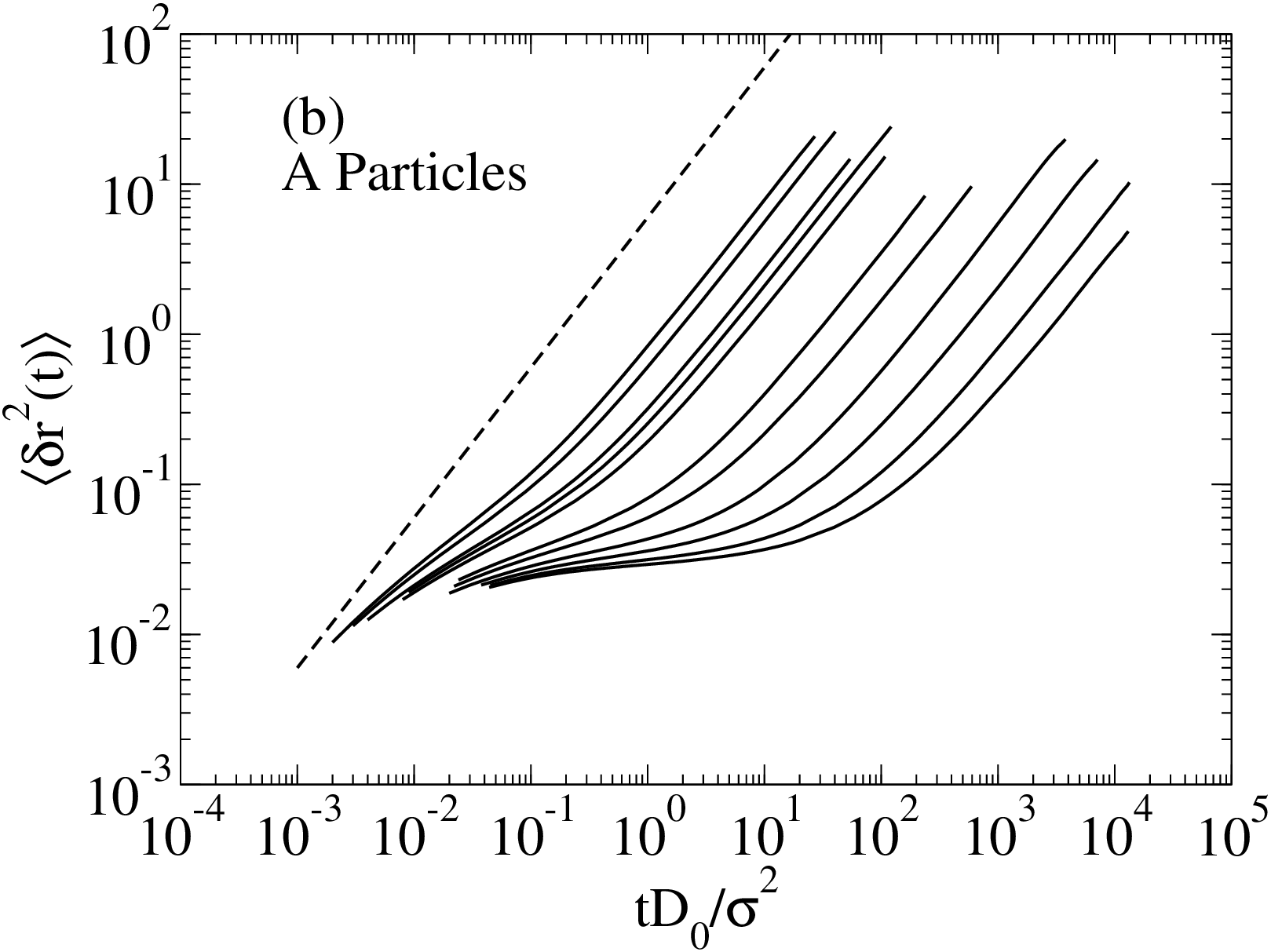}
\caption{\label{msdfig}Mean square displacement for the A particles.
(a) Predicted by the mode-coupling theory.
(b) Calculated from the Brownian dynamics simulations.
The solid lines correspond to the same reduced temperatures 
$\epsilon = (T-T_c)/T_c$ in the
mode-coupling theory calculations and the Brownian dynamics simulations.
The reduced temperatures are
3.5977, 2.4483, 1.2989, 1.0690, 0.8391, 0.3793, 0.2644, 0.1494, 0.0805, 0.0345, 0.0115
listed from left to right.   
The dashed line corresponds to the limit of non-interacting particles and
the dotted line in figure (a) is the mean square displacement calculated at $T_c^{theory}$.}
\end{figure}

We present the mode-coupling theory predictions and simulations results
in Figs.~\ref{msdfig}a and \ref{msdfig}b, respectively, for the A particles. 
The results for the B particles are similar.  
The solid lines correspond to the same reduced temperatures 
for the mode-coupling calculations and the Brownian dynamics
simulations.  The dotted line in Fig.~\ref{msdfig}a is the 
mean square displacement at $T_c^{theory}$. The dashed line in
both figures corresponds to motion in the limit of 
non-interacting particles, \textit{i.e.} a purely diffusive motion with a diffusion 
coefficient $D_0$.  At all temperatures the short time motion is diffusive
with a diffusion coefficient $D_0$.  For the mode-coupling theory calculations,
the short time diffusion coefficient is one, but the short time diffusion
coefficient is temperature dependent in the Brownian dynamics simulations.  
Note, however, that the
time axes are scaled to compensate for this difference.  

The mode-coupling theory correctly predicts the existence of the 
plateau region in the log-log plot of the mean square displacement for
temperature close to the transition temperature.  The plateau
represents a localization of the particles for several decades in time and 
is associated with the cage effect.  According to the mode-coupling theory,
above the transition temperature the long time
motion is diffusive with a self diffusion coefficient $D > 0$. 
Also, at and below 
the mode-coupling transition temperature,
there is structural arrest and $D = 0$.

\begin{figure}
	\includegraphics[scale=0.25]{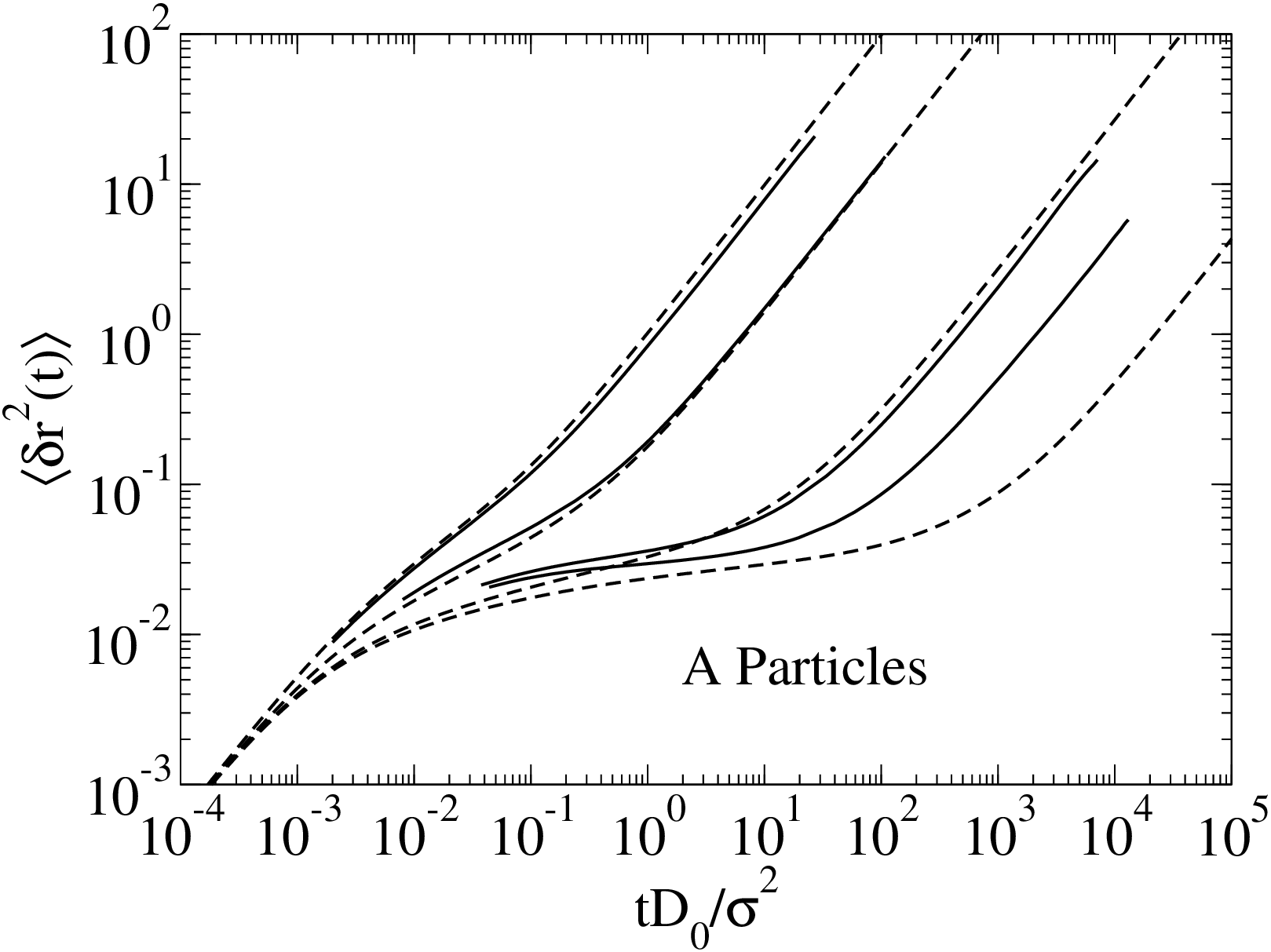}\\[0.25cm]
	\includegraphics[scale=0.25]{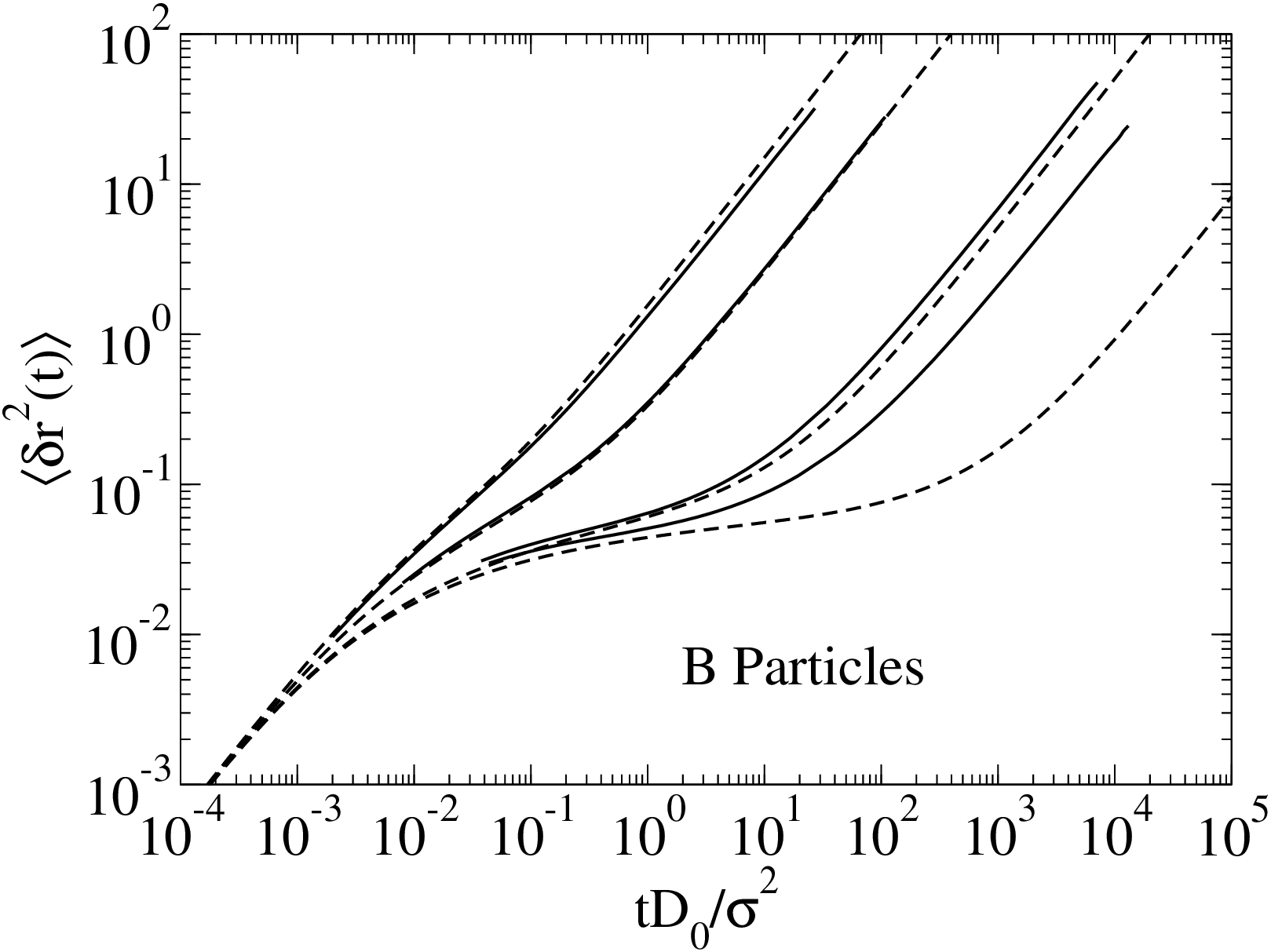}
\caption{\label{msdcomp}The mean square displacement for the A and B particles
predicted by the mode-coupling theory (dashed lines) and calculated
from the Brownian dynamics simulations (solid lines).  The reduced temperatures
are $\epsilon = 3.5977$, 0.839, 0.0805 and 0.0115 listed from
left to right.}
\end{figure}

In Fig.~\ref{msdcomp} we show the mean square displacements
for reduced temperatures $\epsilon = 3.5977$, 0.8391 0.0805, 0.0115
calculated using the mode-coupling theory (dashed lines) and obtained from the
Brownian dynamics simulations (solid lines) for the A and B particles.
The A particles are shown in the upper figure and the B particles are
shown in the lower figure.  
The mean square displacement agrees reasonably
well with the predictions of the mode-coupling theory 
for reduced temperatures
equal to and above 0.0805 
for both the A and B particles at all times. 
However, for reduced temperatures below $0.0805$,
the mean square displacement is not accurately predicted by the
mode-coupling theory.

\begin{figure}
	\includegraphics[scale=0.25]{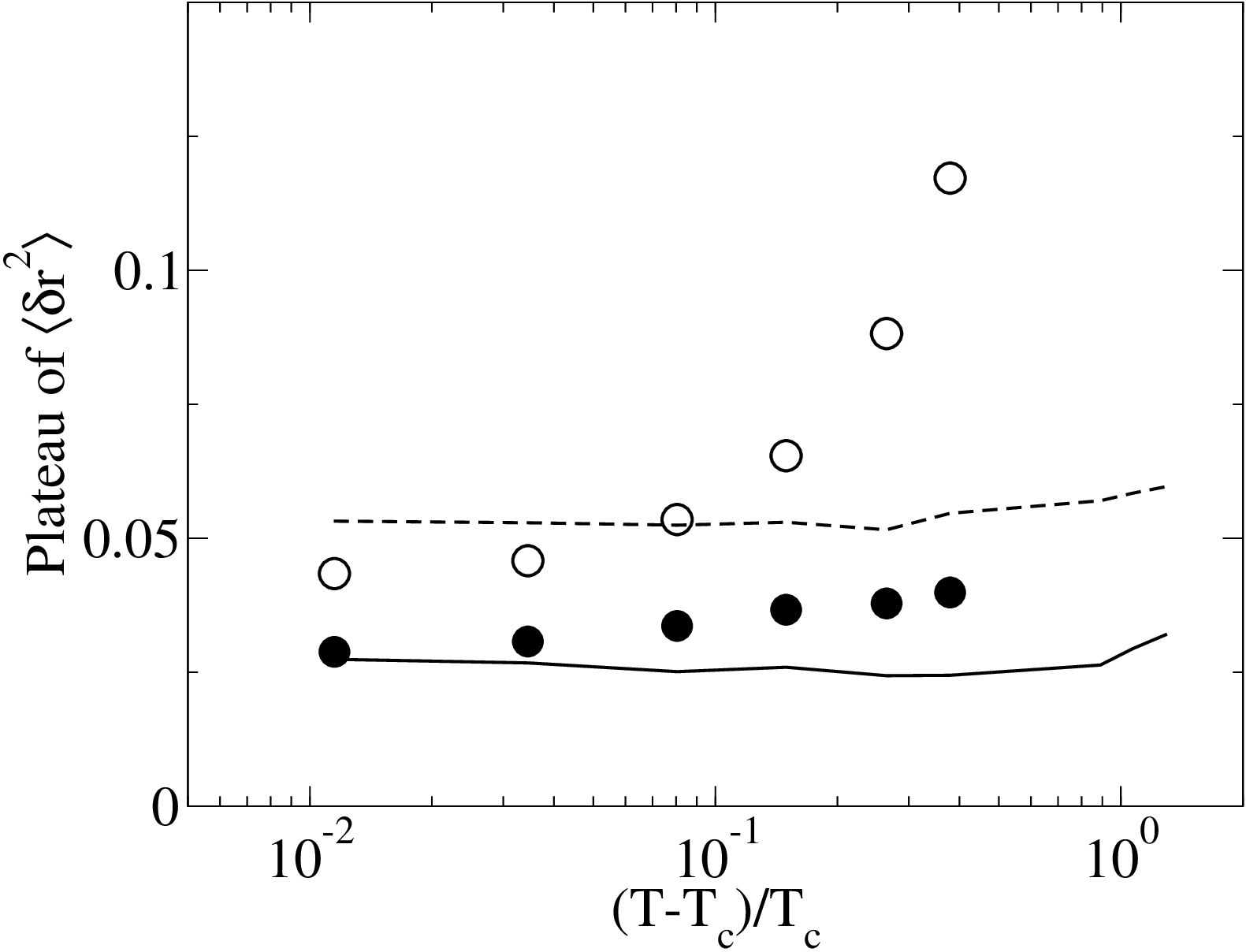}
\caption{\label{plat}Comparison of the plateau value
of the mean square displacement versus temperature.
The symbols are the simulation results and the lines
are the predictions of the mode-coupling theory.  
The A particles are represented by the closed symbols and
the solid line, and the B particles are represented by the open 
symbols and the dashed line.}
\end{figure}

For all reduced temperatures smaller than $\epsilon=0.8391$, there is an obvious
plateau in the Brownian dynamics simulations and in the mode-coupling 
theory calculations. We define the height of the plateau region
as the inflection point in the logarithm of the mean square
displacement versus the logarithm of time.  
At small reduced temperatures the height of the plateau predicted by the theory
agrees reasonably well with that obtained from simulations, Fig.~\ref{plat}.
In particular, at $\epsilon = 0.0115$, the plateau is predicted by the theory at a 
value of the mean square displacement of around 0.028 for the A particles and it occurs
in the Brownian dynamics simulations at around  0.029. 
For the B particles at $\epsilon = 0.0115$, the 
plateau is predicted by the theory to be around  a value of 0.053 and it  
is around 0.043 in the simulations. 
However, the temperature dependence of the plateau 
according to the theory and in the simulations is very different. 
The theory predicts that the plateau height 
as a function of reduced temperature is
essentially constant until around
$\epsilon = 0.8391$, and the plateau height increases slightly with increasing temperature.
The plateau height calculated from the simulations increases with temperature
faster than predicted by the theory. For reduced
temperatures above 0.3793, it is difficult to 
calculate the inflection point for the Brownian dynamics simulations
accurately.  Note that in the temperature range in which the theory gives reasonably
accurate predictions for the incoherent scattering function and the mean square 
displacement, the plateau height resulting from the theory is quite a bit smaller than that
obtained from simulations. For example, for $\epsilon = 0.3793$
the mean square displacement at the inflection point predicted by the theory 
is around 0.024 and 0.053 for the A and B particles,
respectively, whereas in simulations it occurs 
around 0.040 and 0.12  for the A and B particles, 
respectively.  

\begin{figure}
	\includegraphics[scale=0.25]{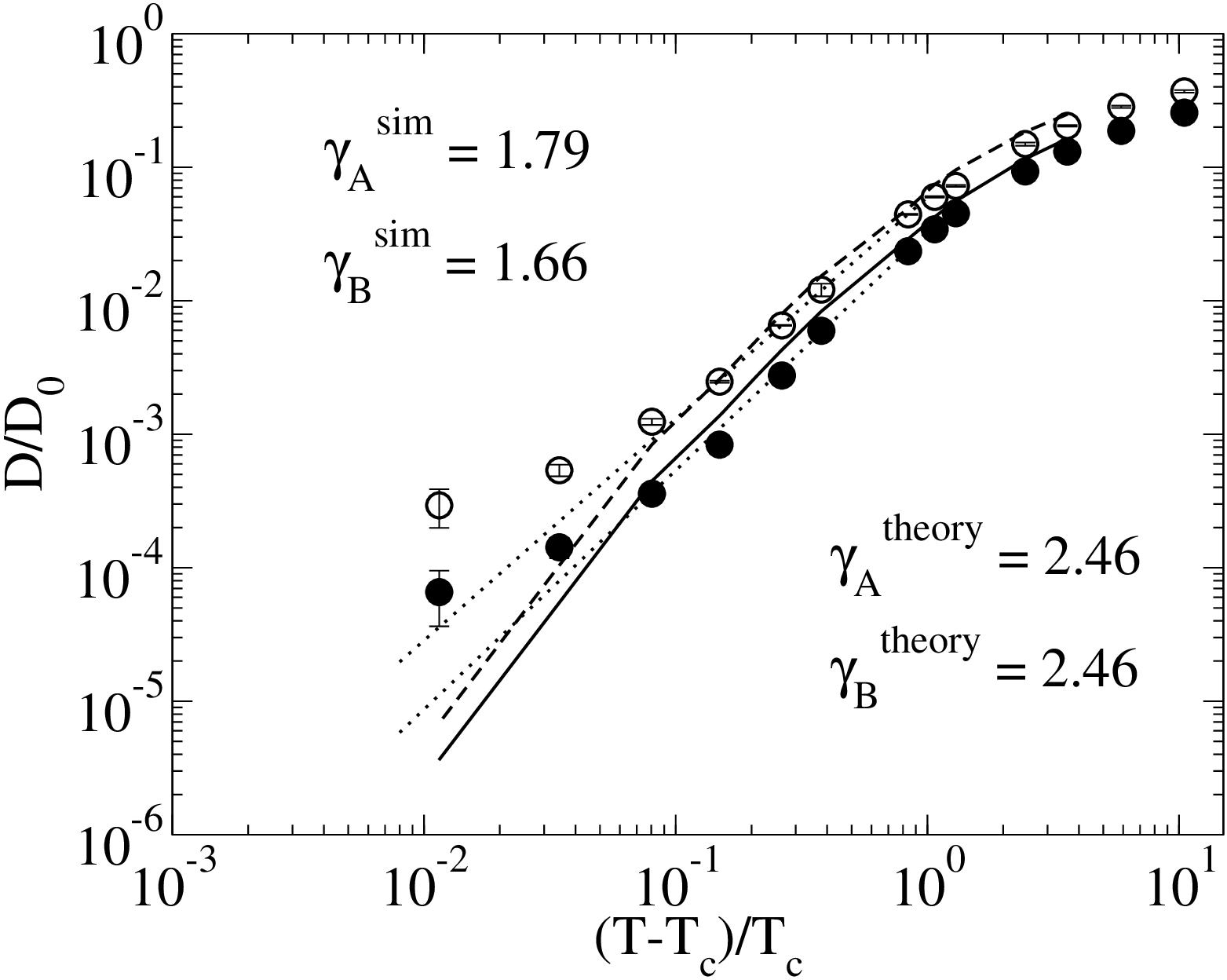}
\caption{\label{diffcomp}The diffusion coefficient 
determined from the Brownian dynamics simulation (symbols)
and the mode-coupling theory (solid and dashed lines).  The closed
symbols and the solid line are the results for the A particles.
The open symbols and the dashed line are the results for the B
particles.  The dotted line is a fit of the
simulation data to the function $a[(T-T_c)/T_c]^{\gamma}$.  The
exponents to the power law fits are given in the figure along
with the exponents to power law fits to the predictions of the
mode-coupling theory.}
\end{figure}

We obtain the self diffusion coefficient 
$D$ from the slope of the mean square displacement at long times, 
\textit{i.e.}~from
$D = \lim_{t \rightarrow \infty} \left< \delta r_{\alpha}^2(t) \right>/(6t)$.
Within the mode-coupling theory, we can also calculate $D$ from the equation 
\begin{equation}
\frac{D}{D_0} = \frac{1}{1 + \int_{0}^{\infty} \mbox{d}t \mathcal{M}_{\alpha}^0(t)}.
\end{equation}
Both procedures agree to within one percent.
The diffusion coefficients predicted by the theory and obtained 
from Brownian dynamics simulations are 
shown in Fig.~\ref{diffcomp} as a function of reduced temperature.  
The mode-coupling theory provides a 
good prediction for the diffusion coefficients for 
$\epsilon \ge 0.0805$, but the diffusion coefficients predicted by the theory are slightly larger
than the ones found from simulations. At lower reduced temperatures the 
theory strongly underestimates
the diffusion coefficients. In addition, the theory does not capture the increasing difference
between the diffusion coefficients of the A and the B particles.

We fit $D/D_0$ to power laws of the form $a[(T-T_c)/T_c]^{\gamma}$; 
the exponents are given in Fig.~\ref{diffcomp}.
For the mode-coupling theory fit we use reduced temperatures less than 0.0345
whereas for the Brownian dynamics simulations we fit the 
range $0.1495 \le \epsilon \le 0.8391$.  
The exponents for the Brownian dynamics simulations are slightly different
than what has been reported in an earlier work \cite{FlennerSzamel2}, since we
are fitting $D/D_0$ here and $D$ there.  
The exponents predicted by the mode-coupling theory are considerably larger than 
those obtained from the fits to simulations' results.  Note that the
exponents for the diffusion coefficient calculated from 
the theory is the same as the exponents found
for the $\alpha$ relaxation time.  Moreover, the exponents 
found from the mode-coupling calculations are
the same for the A and the B particles.

The reduced temperatures in which it is possible to fit
the mode-coupling results well with a power law
is similar to what was found for the 
$\alpha$ relaxation time, \textit{i.e.} that the 
power law provides a good fit up to a 
reduced temperature of around 0.08.  

We would like to point out that power laws fit the predictions of the
mode-coupling theory reasonably well for
the same reduced temperatures in which  
we fit power laws to the results of the Brownian dynamics simulations.
If we fit the predictions of the mode-coupling theory to power laws using the range 
$0.1495 \le \epsilon \le 0.8391$, the resulting exponents are different from
the true exponents (\textit{i.e.} from the exponents describing the the true
asymptotic power law behavior) but they differ
by at most 12\% from the exponents obtained from the
Brownian dynamics simulations. This should be compared with 
the 48\% difference between the true mode-coupling theory exponent 
and the exponent for the B particles 
obtained from the Brownian dynamics simulations using the range 
$0.1495 \le \epsilon \le 0.8391$.

\begin{figure}
	\includegraphics[scale=0.25]{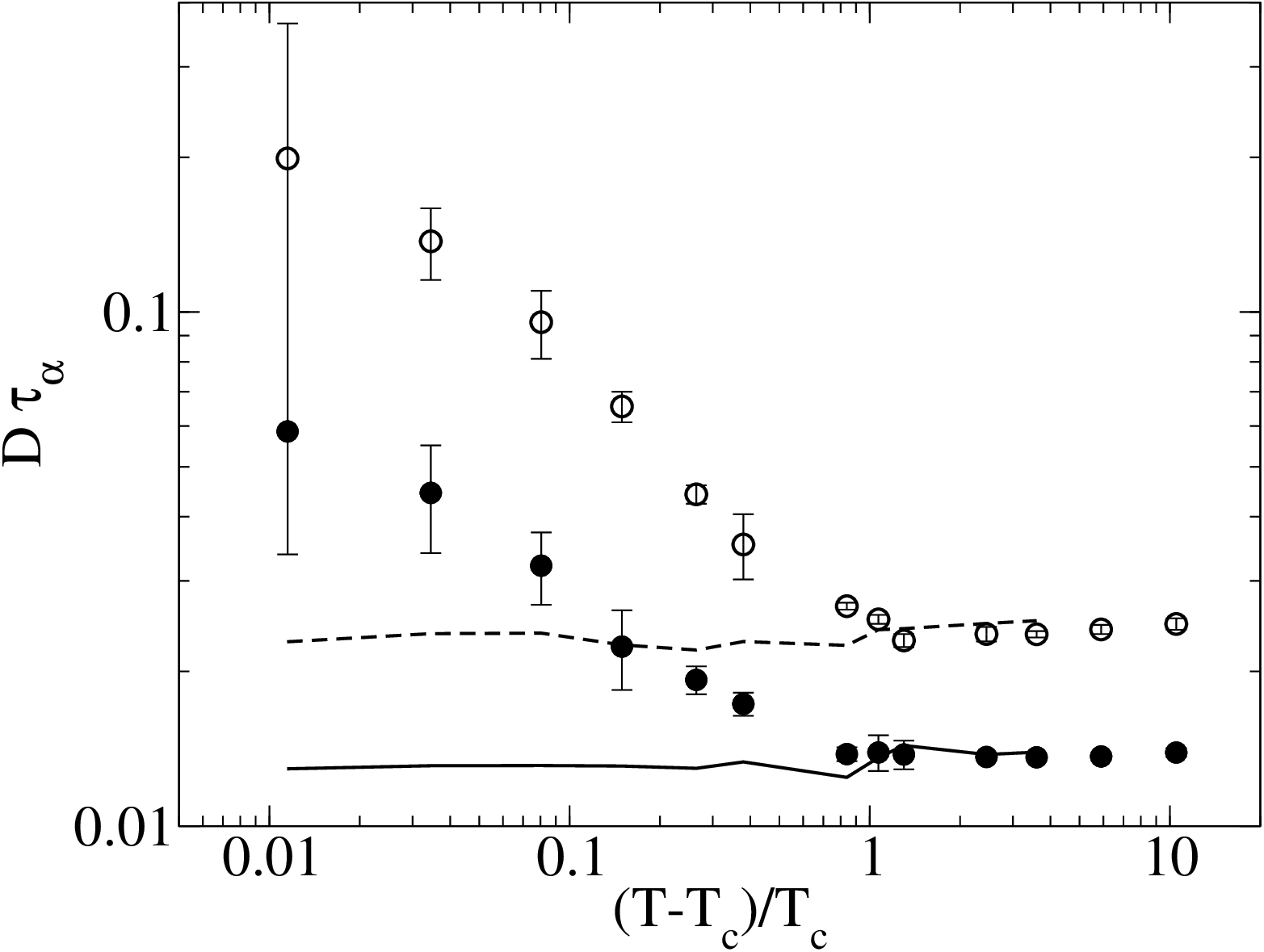}
\caption{\label{difftau} Product of the diffusion coefficient and the
$\alpha$ relaxation time 
determined from the Brownian dynamics simulation (symbols)
and the mode-coupling theory (solid and dashed lines).  The closed
symbols and the solid line are the results for the A particles.
The open symbols and the dashed line are the results for the B
particles.  Note that for clarity we omitted the error bars for the 
lowest temperature point ($\epsilon = 0.0115$) for the A particles;
at the lowest temperature for the A particles $D\tau_{\alpha} = 0.0585 \pm 0.054$}
\end{figure}

Finally, in Fig. \ref{difftau} we compare the temperature dependence of the 
product of the diffusion coefficient and the
$\alpha$ relaxation time predicted by the mode-coupling theory and obtained 
from simulations. In the high temperature regime one usually expects that the 
Stokes-Einstein relation is valid and $D\tau_{\alpha}$ is temperature-independent.
The decoupling of the diffusion and structural relaxation
has been identified as one of the signatures of increasingly heterogeneous 
dynamics \cite{MEreview}. The mode-coupling theory predicts an essentially
temperature-independent $D\tau_{\alpha}$. In contrast, simulations show that
the product of the diffusion coefficient and the
$\alpha$ relaxation time starts increasing with temperature below approximately
$\epsilon=1$. Note that this value of the reduced temperature 
corresponds to a temperature that is 
close to the so-called onset
temperature identified for the Kob-Andersen model by Brumer and Reichman \cite{onset}.

\section{\label{nongauss} Non-Gaussian Parameter}

At short and long times, the motion of the particles is 
Fickian and the self part of the
van Hove correlation function is Gaussian.
For intermediate times, the van Hove correlation function
deviates from Gaussian.
To examine the non-Gaussian nature of the
van Hove correlation function, we 
calculated the non-Gaussian parameter 
\begin{equation}
\alpha_2(t) = \frac{3}{5}\frac{\left<\delta r^4_{\alpha}(t)\right>}
{\left<\delta r^2_{\alpha}(t)\right>^2} - 1.
\label{ngeq}
\end{equation}
For the calculation of $\alpha_2(t)$, we first calculated the 
mean square displacement (see section \ref{msd}) and then we calculated 
$\left< \delta r^4_{\alpha}(t) \right>$.
The equation of motion for $\left< \delta r^4_{\alpha}(t) \right>$ is derived 
using the same method as the equation
for $\left< \delta r^2_{\alpha}(t) \right>$.  The 
resulting equation of motion is
\begin{eqnarray}
\frac{\partial}{\partial t} \left< \delta r^4_{\alpha}(t) \right> & = & 20 D_0 
\left< \delta r^2_{\alpha}(t) \right> \nonumber \\
& & - \int_{0}^t \mbox{d}u \mathcal{M}_{\alpha}^0(t-u) 
\frac{\partial}{\partial u} \left< \delta r^4_{\alpha}(u) \right> 
\nonumber \\
& & + 10 \int_0^t \mbox{d}u \mathcal{M}_{\alpha}^2(t-u) 
\frac{\partial}{\partial u} \left< \delta r^2_{\alpha}(u) \right> \nonumber \\
\label{r4eq}
\end{eqnarray}
where
\begin{eqnarray}
\mathcal{M}_{\alpha}^2(t) & = &\frac{V D_0^2}{10 \pi^2 N_{\alpha}} \nonumber \\ 
& & \times \int \mbox{d}k\ k^4 \left[ \frac{2}{3 k} 
\frac{\partial}{\partial k} F_{\alpha}^s(k,t) + 
\frac{\partial^2}{\partial k^2} F_{\alpha}^s(k,t) \right] \nonumber \\
& & \times 
\sum_{\delta \delta^{\prime}}C_{\alpha \delta}(k) 
S_{\delta \delta^{\prime}}(k,t)C_{\alpha \delta^{\prime}}(k).
\end{eqnarray}
In principle, it is also possible to calculate the non-Gaussian parameter
by fitting $F_{\alpha}^s(q,t)$ in the small wave vector limit \cite{Kaur}.  We
found that this procedure was difficult to follow using the 
structure factors calculated from simulations, and small
numerical uncertainties in the small $q$ values of $F_{\alpha}^s$ 
can result in large changes in the non-Gaussian parameter.
For short times, the integrals involving memory functions 
in Eqs.~(\ref{msdeq}) and (\ref{r4eq}) are 
close to zero.  
Thus, for short times $\left< \delta r^2(t) \right> \approx 6 D_0 t$ and 
$\left< \delta r^4(t) \right> \approx 60 D_0^2 t^2$, therefore $\alpha_2(t) \approx 0$.  
At long times, the
motion is once again Fickian, the self part of the
van Hove correlation function
is Gaussian and $\alpha_2(t) = 0$.  In our calculations,
$\alpha_2(t)$ does not go to zero at long times, 
but rather to a value around $5\times10^{-3}$.  We believe that this result can be attributed
to numerical errors; it does not imply that $\alpha_2(t)$ is
nonzero for $t \rightarrow \infty$.
  
\begin{figure}
	\includegraphics[scale=0.25]{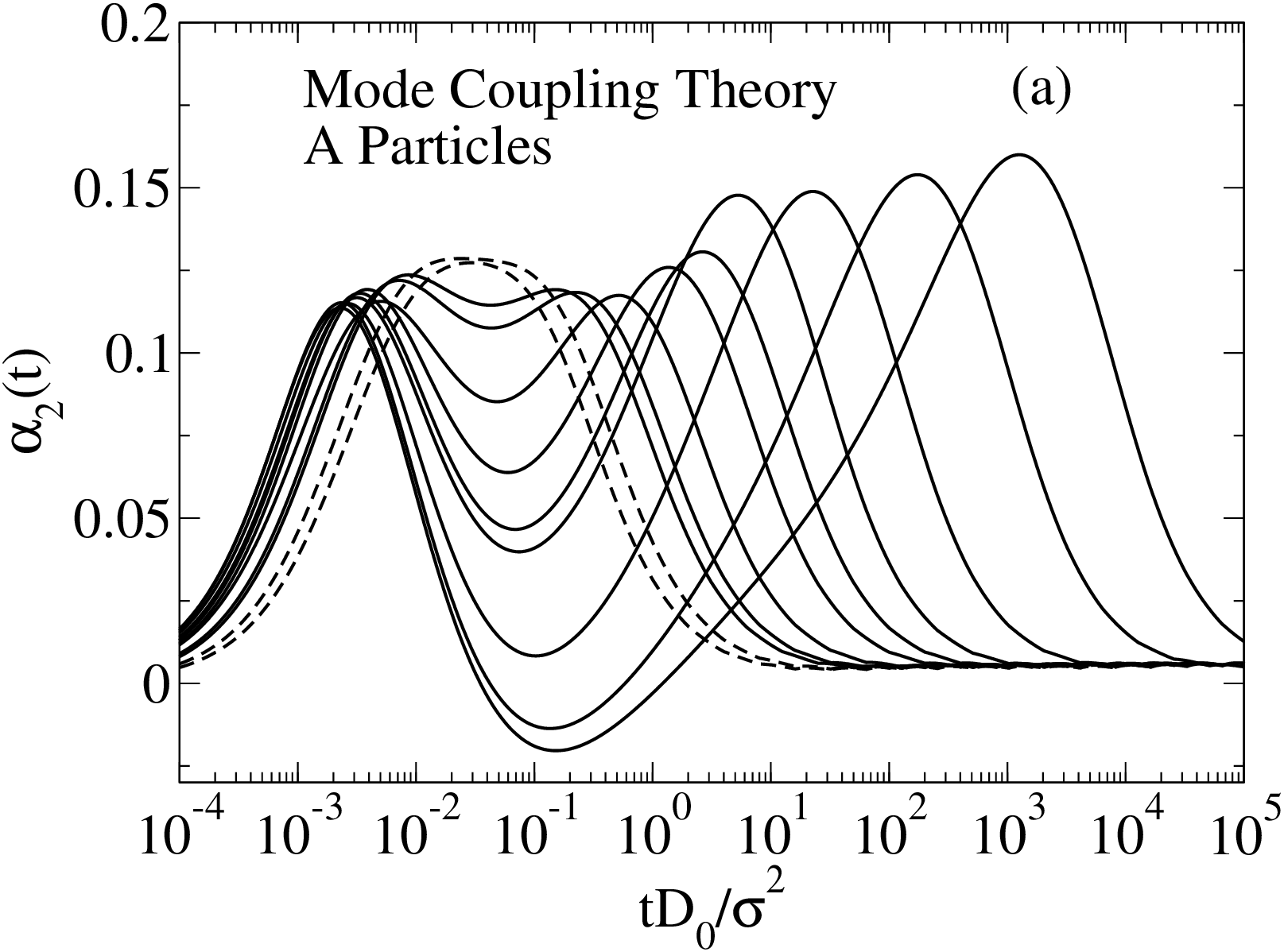}\\[0.25cm]
	\includegraphics[scale=0.25]{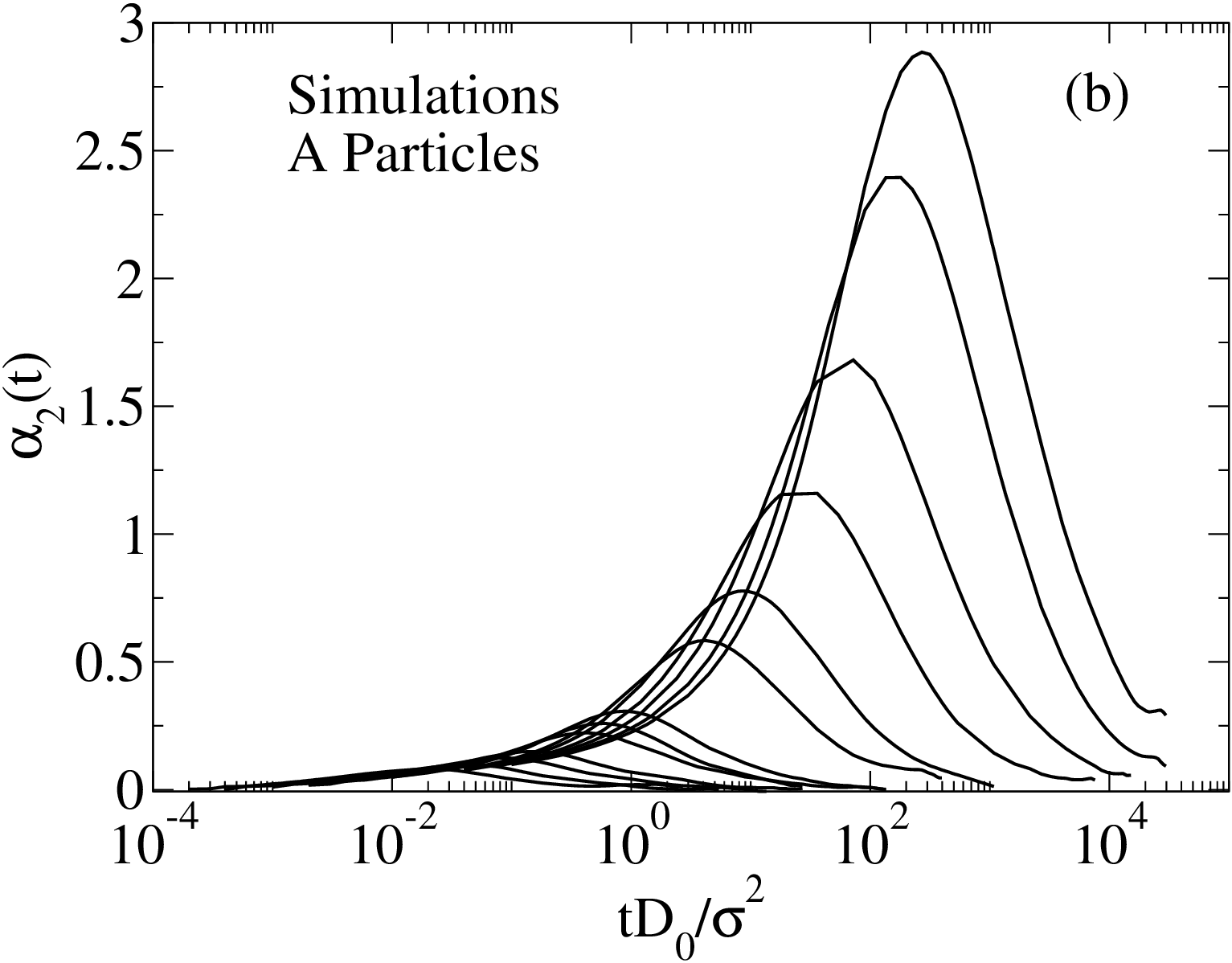}
\caption{\label{ngfiga}The non-Gaussian parameter for the A
particles. 
(a) Predicted by the mode-coupling theory.
(b) Calculated using the Brownian dynamics simulations.
The reduced temperatures $\epsilon = (T-T_c)/T_c$ are
3.5977, 2.4483, 1.2989, 1.0690, 0.8391, 0.3793, 0.2644, 0.1494, 0.0805, 0.0345, 0.0115.  
In figure (a), there is one peak
for $\epsilon = 3.5977$, and  a wider single peak for 2.4483 (dashed lines).
For all the other reduced temperatures there are two peaks, and the
larger peak position of the second peak corresponds to a 
lower reduced temperature.  In figure (b), the larger peak
heights correspond to lower reduced temperatures.}
\end{figure}
\begin{figure}
	\includegraphics[scale=0.25]{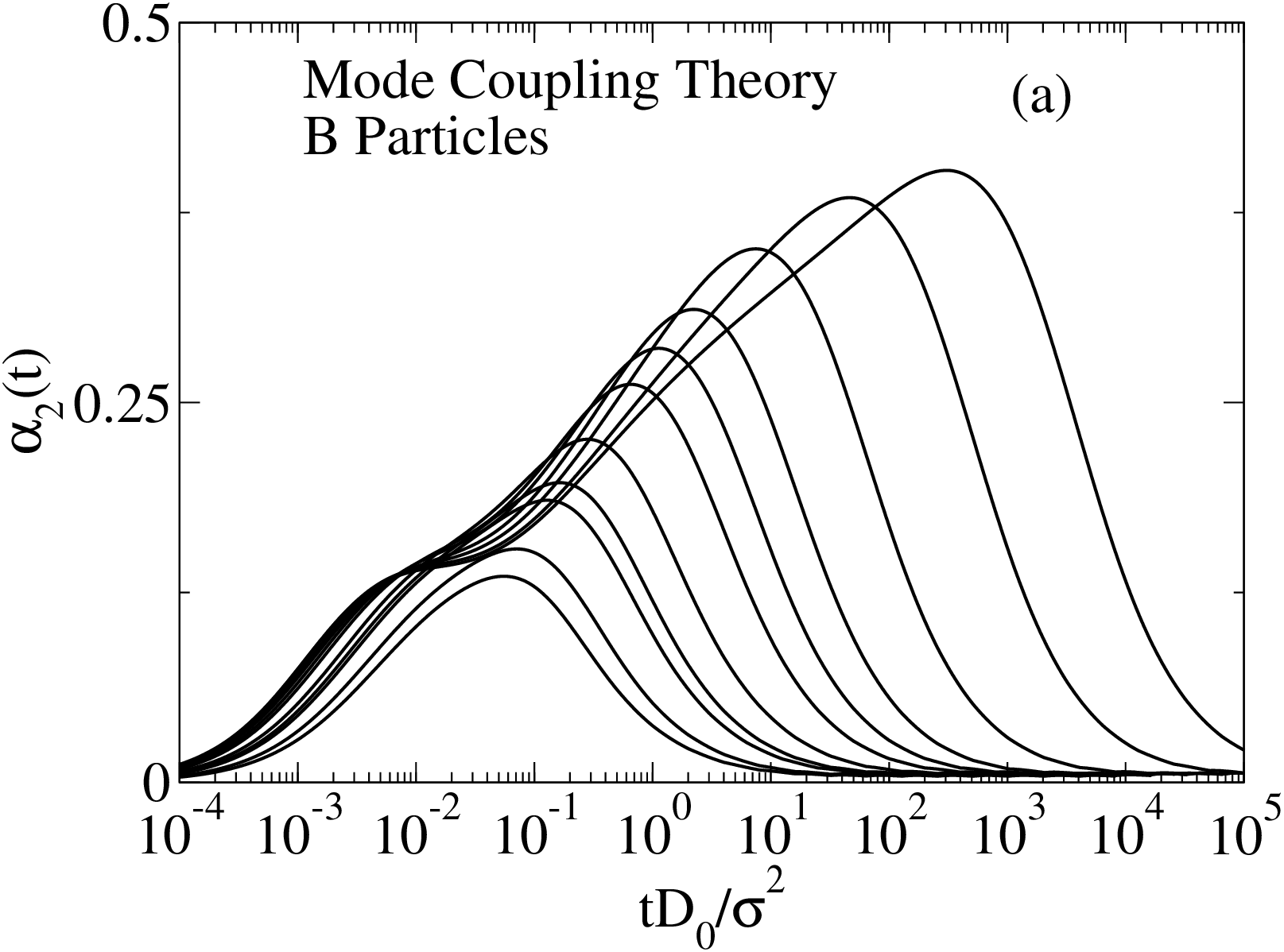}\\[0.25cm]
	\includegraphics[scale=0.25]{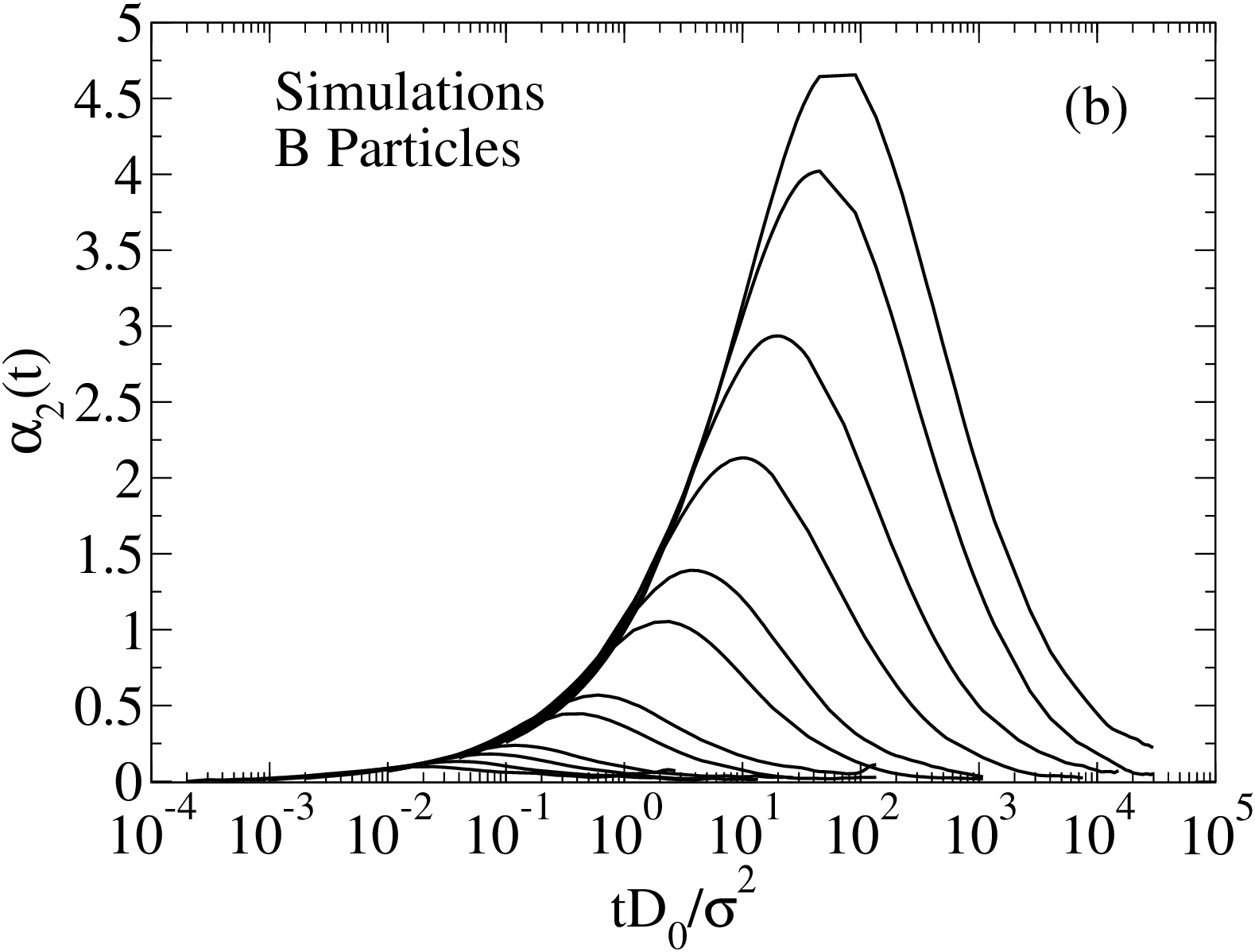}
\caption{\label{ngfigb}The non-Gaussian parameter for the
B particles.
(a) Predicted by the mode-coupling theory.
(b) Calculated from the Brownian dynamics simulations.
The reduced temperatures $\epsilon = (T-T_c)/T_c$ are
3.5977, 2.4483, 1.2989, 1.0690, 0.8391, 0.3793, 0.2644, 0.1494, 0.0805, 0.0345, 0.0115
where the lower reduced temperatures
correspond to larger peak heights.}
\end{figure}

\begin{figure}
	\includegraphics[scale=0.25]{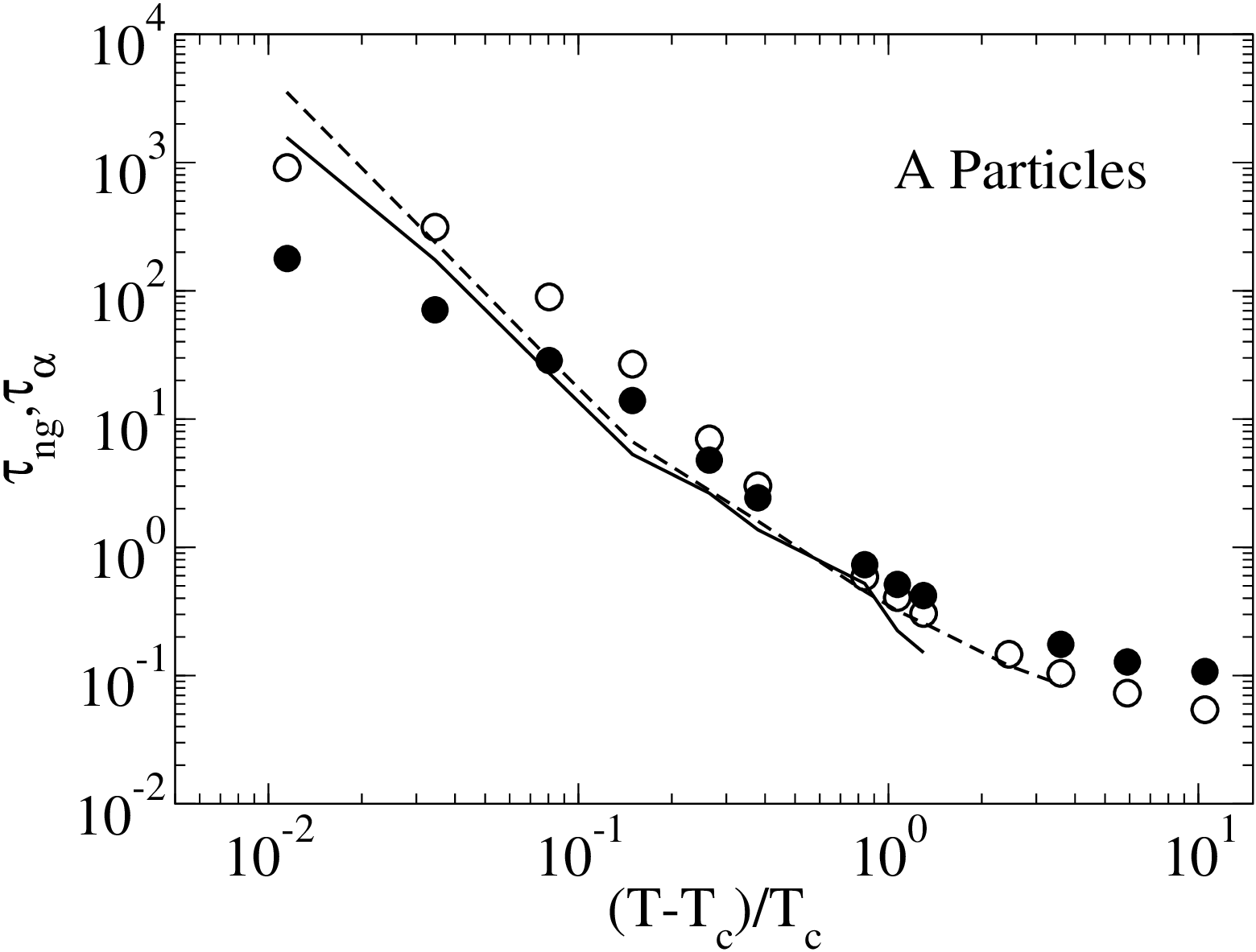}\\[0.25cm]
	\includegraphics[scale=0.25]{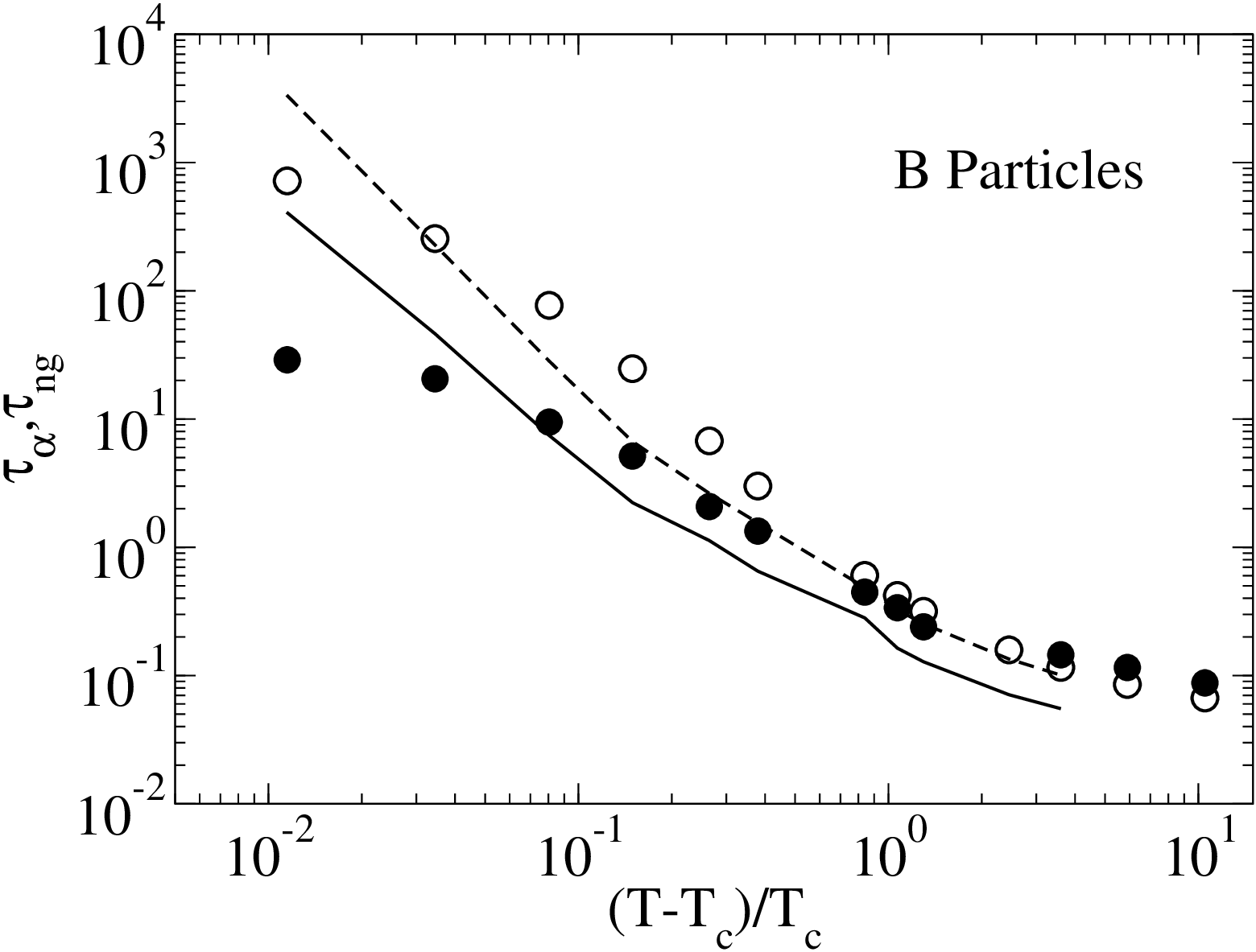}
\caption{\label{peakcomp}The peak position of the non-Gaussian
parameter compared to the $\alpha$ relaxation time.  The
symbols are the simulation results and the lines are the predictions
of the mode-coupling theory.  The open symbols and dashed lines are
the $\alpha$ relaxation time.  The closed symbols and the solid lines
are the peak positions of the non-Gaussian parameter.}
\end{figure}

We show the non-Gaussian parameters for the A particles in Fig.~\ref{ngfiga}. 
The upper panel shows the predictions of 
the mode-coupling theory and the lower panel shows the results of  
the Brownian dynamics simulations.  
According to the mode-coupling calculations, there is one peak at high reduced 
temperatures, but for $\epsilon \le 1.2989$ there are two peaks. Two peaks have
been observed before in mode-coupling calculations 
of $\alpha_2(t)$ for other systems \cite{Kaur}. 
The first peak is around the beginning of the 
plateau region of the mean square displacement and 
the initial decay of the scattering functions.  The 
position of the first peak decreases for decreasing temperature, but
is almost constant for $\epsilon \le 0.8391$.  
The second peak is around the $\alpha$ relaxation time which corresponds to
just after the plateau region of the mean square displacement. 
The position of the second peak increases with decreasing temperature and 
roughly follows the temperature dependence of the $\alpha$ relaxation
time, Fig.~\ref{peakcomp}a.    

There is only one peak in $\alpha_2(t)$ according to 
the Brownian dynamics simulations, Fig.~\ref{ngfiga}b. The position
of this peak is greater than the $\alpha$ relaxation time at 
higher temperatures, but it increases slower with
decreasing temperature than the $\alpha$ relaxation time
starting at around $\epsilon = 0.8391$.  

The mode coupling theory predicts one peak at all temperature 
for the B particles.  However, there is a prominent shoulder for
$\epsilon \le 1.2989$, the same reduced temperatures in which there
are two peaks in the non-Gaussian parameter for the 
A particles.  The position of the shoulder follows the same temperature
dependence as the position of the first peak in the non-Gaussian 
parameter for the A particles.  The peak position 
in $\alpha_2(t)$ predicted by the mode-coupling theory 
is less than the $\alpha$ relaxation time at 
all temperatures, and it increases with decreasing temperature at close
to the same rate as the $\alpha$ relaxation time, Fig.~\ref{peakcomp}b.

The non-Gaussian parameter for the B particles obtained from
the Brownian dynamics simulation is shown in Fig.~\ref{ngfigb}b.
There is only one peak for all temperatures.  The position of
the peak is greater than the $\alpha$ relaxation time for 
higher temperatures, but the position increases slower with
decreasing temperature than the $\alpha$ relaxation time 
and is much less than the $\alpha$ relaxation time
at small reduced temperatures, Fig.~\ref{peakcomp}b.

To summarize, the time dependence of 
the non-Gaussian parameter calculated using the mode-coupling
theory is significantly different from what is obtained from 
the Brownian dynamics simulations.  
More importantly, the mode-coupling theory strongly underestimates the deviations from 
Gaussian (\textit{i.e.} Fickian) diffusive motion: 
the heights of non-Gaussian parameters predicted by the theory are almost an order of
magnitude smaller than those obtained from the Brownian dynamics simulations. We show in the 
next section that the simulations show even stronger non-Gaussian effects on somewhat longer 
time scales. 

\section{\label{vanhove} Probability distributions of the logarithm of 
single particle displacements}

In an earlier work \cite{FlennerSzamel2}, we showed that as the 
temperature $T_c^{sim} = 0.435$ is approached, the motion of the 
particles changes from a high temperature diffusive-like behavior to 
a low temperature hopping-like motion. 
To this end, we investigated 
probability distributions
of the logarithm of single particle displacements \cite{Reichman,Puertas}.
The probability distribution of the logarithm 
of single particle displacements 
at a time t, $P\bm{(}\log_{10}(\delta r);t\bm{)}$,  
can be obtained from the self van Hove correlation
function, 
$G_s(\delta r,t) = \left<\delta\left(|\vec{r}\,_i^{\alpha}(t)-
\vec{r}\,_i^{\alpha}(0)| - \delta r\right)\right>$,
by the following transformation 
$P\bm{(}\log_{10}(\delta r);t\bm{)} = \ln(10) 4 \pi \delta r^3 G_s(\delta r,t)$.  
Note that if the motion of a tagged particle is diffusive at all
times with a diffusion coefficient $D$, then the self
van Hove correlation function 
$G_s(\delta r,t) = (1/(4 \pi D t)^{\frac{3}{2}}) \exp(-\delta r^2/4 D t)$ and 
it can be shown that the shape of the probability distribution 
$P\bm{(}\log_{10}(\delta r);t\bm{)}$
is time-independent. In particular, 
the peak height of $P\bm{(}\log_{10}(\delta r);t\bm{)}$ does not depend time and is equal to 
$\ln(10) \sqrt{54/ \pi}\ e^{-3/2} \approx 2.13$. Thus,
deviations from this height represent deviations from Gaussian behavior of 
$G_s(\delta r,t)$. 

\begin{figure}
	\includegraphics[scale=0.25]{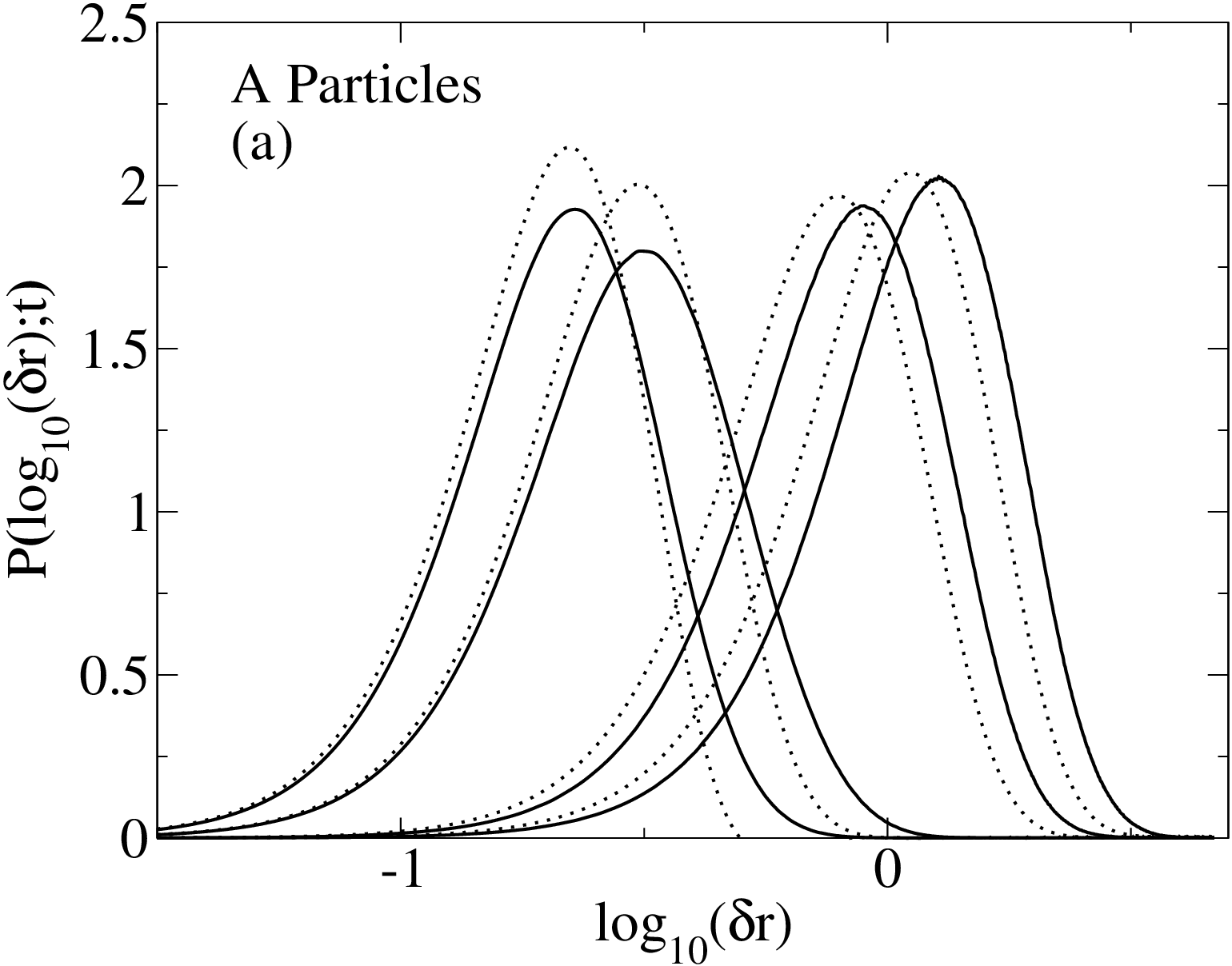}\\[0.25cm]
	\includegraphics[scale=0.25]{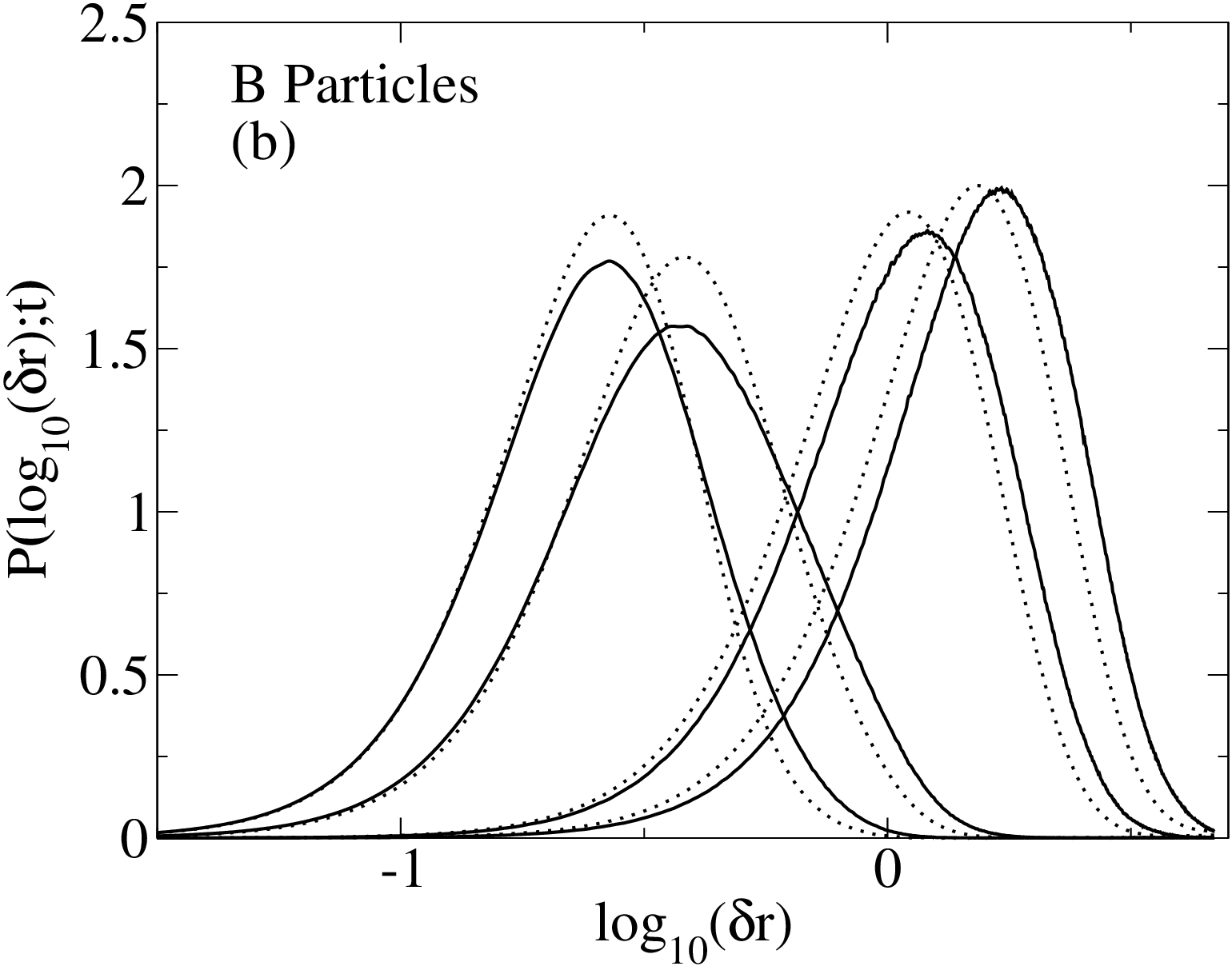}
\caption{\label{plog1}The probability distribution of the logarithm of 
single particle displacements predicted by the mode-coupling theory (dotted lines) 
compared to the probability distributions calculated 
from the Brownian dynamics simulations (solid lines) for $\epsilon = 0.8391$.
(a) A particles.
(b) B particles.
The times shown are 0.25, 1.0, 10, and 20 times the $\alpha$ relaxation time;
note that we use the A-particles and the B-particles $\alpha$ relaxation time in (a) and (b),
respectively.}
\end{figure}

\begin{figure}
	\includegraphics[scale=0.25]{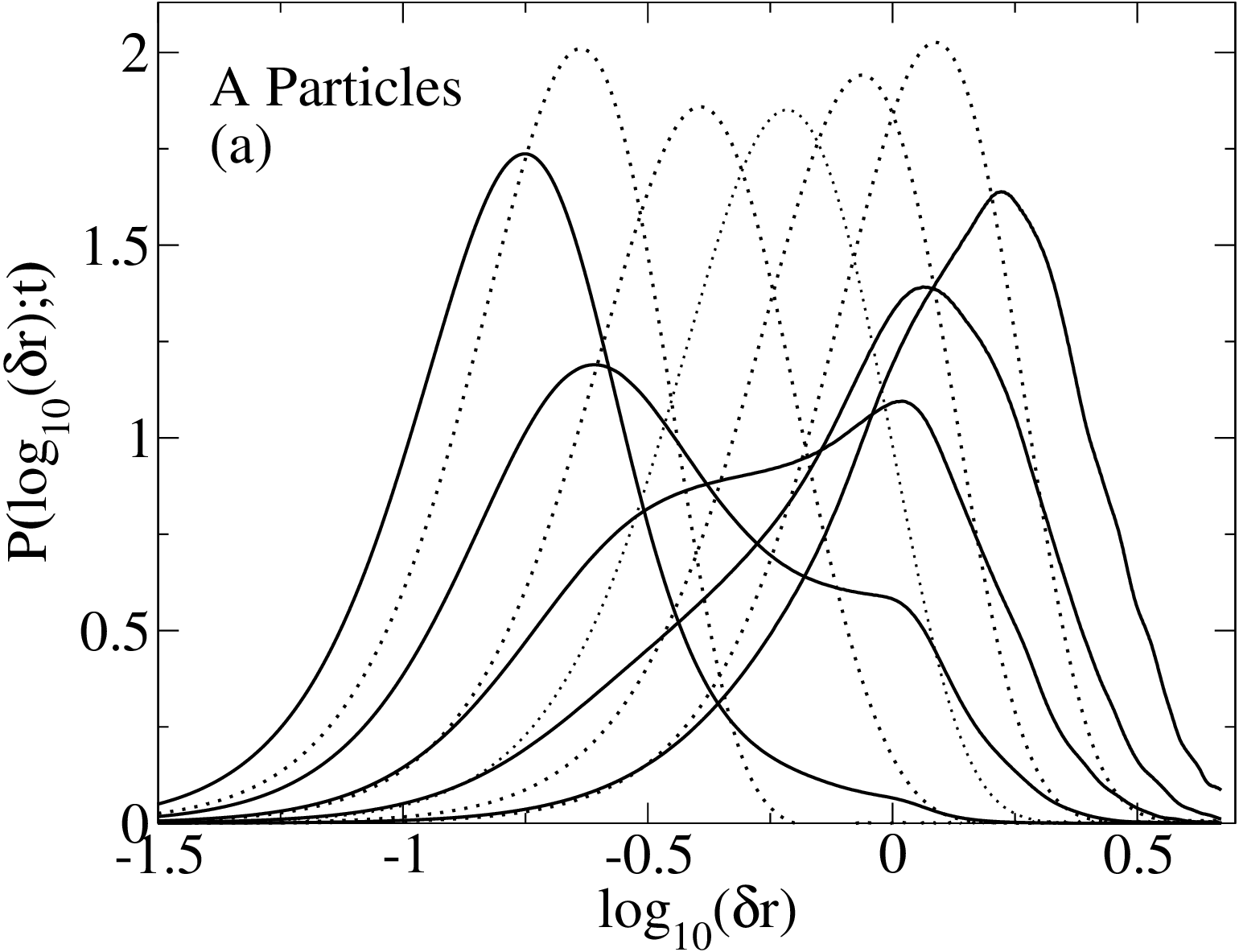}\\[0.25cm]
	\includegraphics[scale=0.25]{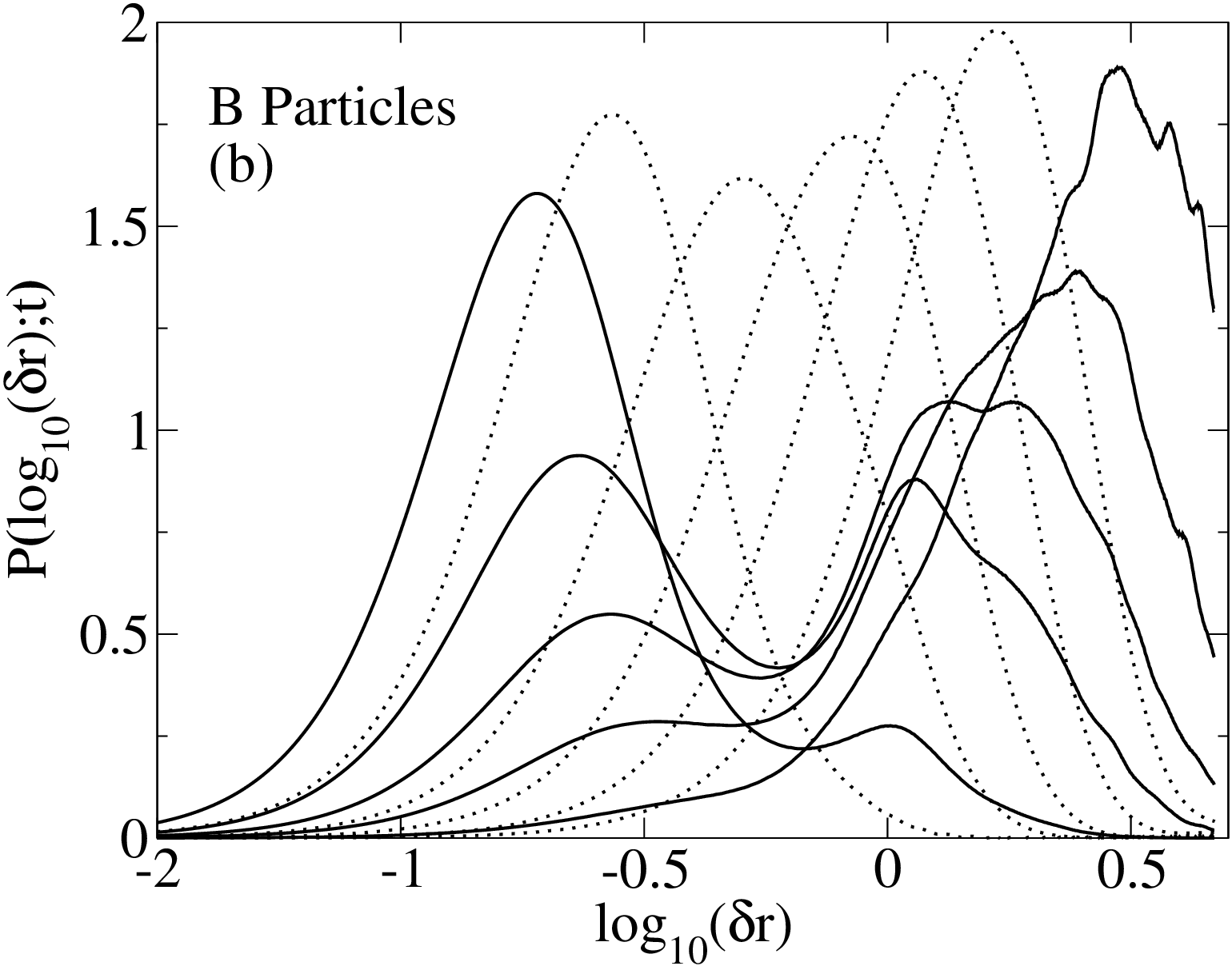}
\caption{\label{plog2}The probability distribution of the logarithm of 
single particle displacements predicted by the mode-coupling theory (dotted lines) 
compared to the probability distributions calculated 
from the Brownian dynamics simulations (solid lines) for $\epsilon = 0.0115$.
(a) A particles.
(b) B particles.
The times shown are 0.1, 1, 2.5, 5, and 10 times the $\alpha$ relaxation
time; note that we use the A-particles and the B-particles $\alpha$ relaxation time in (a) 
and (b), respectively.}
\end{figure}

We show a comparison of 
$P\bm{(}\log_{10}(\delta r);t\bm{)}$ in Fig.~\ref{plog1} for the A and B particles
calculated from the simulations (solid lines)
and from the mode-coupling theory (dotted lines) for the 
reduced temperature $\epsilon = 0.8391$ for several different times:
0.25, 1.0, 5.0, and 10.0 $\tau_{\alpha}$. It should be noted that in Figs.~\ref{plog1}
and \ref{plog2} we use the A and the B particles $\alpha$ relaxation times in panels (a) and (b), 
respectively; moreover, we use
the $\alpha$ relaxation times obtained from simulations for the simulation results
and the $\alpha$ relaxation times predicted by the mode-coupling theory for the
theoretical results. 
At this temperature the mode-coupling theory describes 
the probability distributions reasonably well.
 
\begin{figure}
	\includegraphics[scale=0.25]{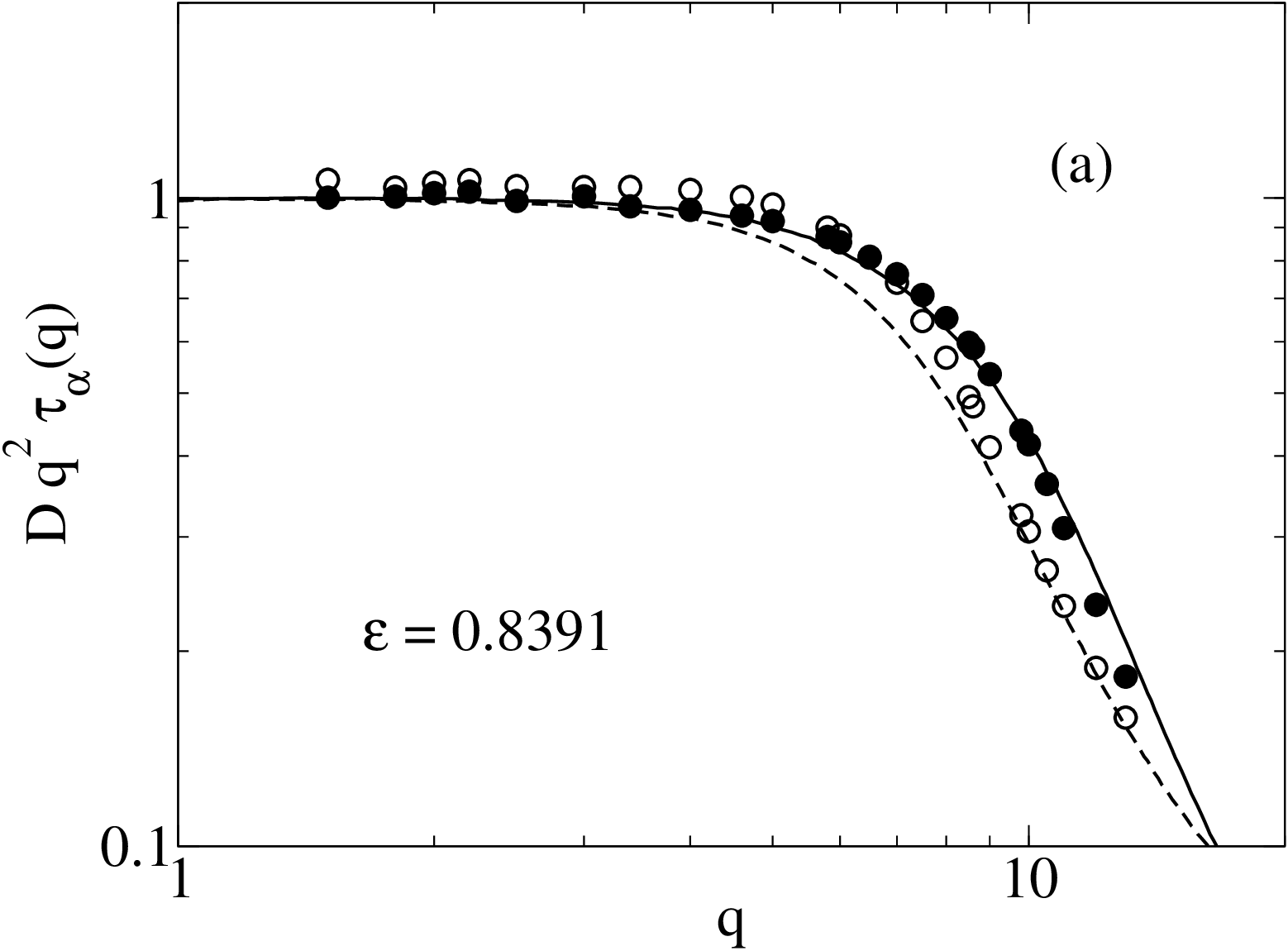}\\[0.25cm]
	\includegraphics[scale=0.25]{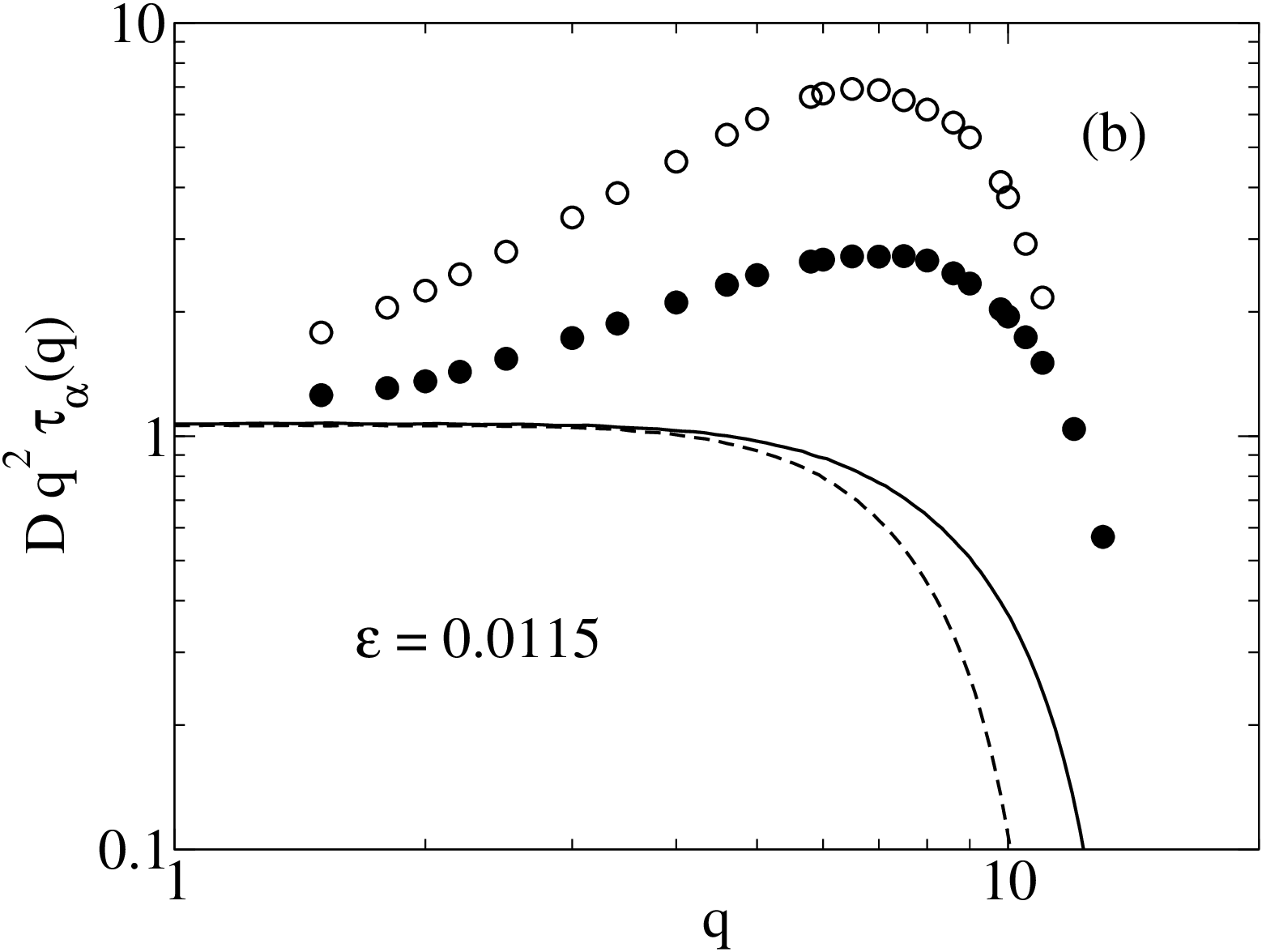}
\caption{\label{ktau}The product $D q^2 \tau_{\alpha}(q)$ 
for $\epsilon = 0.8391$ and 0.0115.  The symbols are the results of the
Brownian dynamics simulations and the lines are the predictions of
the mode-coupling theory.  The A particles are represented by the
closed symbols and the solid lines, and the B particles are represented by
the open symbols and the dashed lines.}
\end{figure}

In Fig.~\ref{plog2} we show a comparison of 
$P\bm{(}\log_{10}(\delta r);t\bm{)}$ for the A and B particles
calculated from the simulations (solid lines)
and from the mode-coupling theory (dotted lines) for
several different times at a reduced temperature
of $\epsilon = 0.0115$ (recall that, as in Fig.~\ref{plog1}, we use
the A and the B particles $\alpha$ relaxation times in panels (a) and (b), 
respectively; moreover we use
the $\alpha$ relaxation times obtained from simulations for the simulation results
and the $\alpha$ relaxation times predicted by the mode-coupling theory for the
theoretical results).  This is the lowest 
reduced temperature in which we can directly compare the
predictions of the mode-coupling theory to the simulation
results.  There is a
dramatic difference in the shape of the curves over the time
interval shown in the figure.  At intermediate times bimodal distributions are obtained 
from Brownian dynamics simulations whereas the mode-coupling theory
predicts unimodal distributions at all times. The bimodal distributions
suggests that a portion of the particles are undergoing
hopping-like motion with a large distribution of 
hopping rates. This hopping-like motion is not predicted by the
mode-coupling theory.

In an earlier work \cite{FlennerSzamel2}, 
we observed in the simulations that the time in which both peaks are about equal
height is longer than the time indicated by the peak position
of the non-Gaussian parameter $\alpha_2(t)$.  We defined a new
parameter 
$\gamma(t) = \frac{1}{3} \left< \delta r^2 \right> \left< 1/\delta r^2 \right> -1$
whose peak position occurs at the same time as when the two 
peaks in $P\bm{(}\log_{10}(\delta r);t\bm{)}$ are approximately the
same height for temperatures in which there are two peaks.  
It was found that the peak position of $\gamma(t)$ has the same temperature dependence as the
$\alpha$ relaxation time.  

At temperatures in which we see evidence of the
hopping-like motion when we examine the probability
distribution $P\bm{(}\log_{10}(\delta r);t\bm{)}$, the
wave vector dependent $\alpha$ relaxation time is
qualitatively different in the simulation than 
predicted by mode-coupling theory. 

In Fig.~\ref{ktau} we show $D q^2 \tau_{\alpha}(q)$,
where $\tau_{\alpha}(q)$ is the wave vector dependent
$\alpha$ relaxation time defined as the
time when $F_{\alpha}^{s}\bm{(}q,\tau_{\alpha}(q)\bm{)} = e^{-1}$.  
The lines are the predictions of the mode-coupling theory and the
circles are the results of the Brownian dynamics simulations.
For small wave vectors, the product $D q^2 \tau_{\alpha}(q)$
is one.  
For the reduced temperature of $\epsilon = 0.8391$,
Fig.~\ref{ktau}a, 
$\tau_{\alpha}(q)^{-1} = D q^2$ until $q \approx 5$. 
At larger wave vectors there is a crossover from the small wave vector relationship
to the large wave vector limit $\tau_{\alpha}(q)^{-1} = D_0 q^2$
(this large wave vector limit  follows from the
fact that memory functions vanish in the large wave vector limit). 

In contrast, for lower reduced temperatures, there is a qualitative difference between 
the wave vector dependent $\alpha$ relaxation
time predicted by the theory and calculated from the Brownian dynamics simulations.  
In Fig.~\ref{ktau}b we show the product 
$D q^2 \tau_{\alpha}(q)$ for the reduced 
temperature $\epsilon = 0.0115$.  At small wave vectors the simulation results approach
the asymptotic behavior  
$\tau_{\alpha}(q)^{-1} = Dq^2$.  At intermediate wave vectors, $1.5 \le q \le 7.5$, 
the values of $D q^2 \tau_{\alpha}(q)$ calculated from the simulation
increase with a peak somewhere
between $6.5 < q < 7.5$.  At large wave vectors, the $\alpha$ relaxation time reaches 
its limiting behavior
$\tau_{\alpha}(q)^{-1} = D_0 q^2$.  On the other hand, the mode
coupling theory predicts a monotonically decaying
$D q^2 \tau_{\alpha}(q)$ with essentially the same behavior at low and high reduced temperatures, 
Fig.\ref{ktau}. 

We should point out that the increase of $D q^2 \tau_{\alpha}(q)$ 
with wave vector for intermediate wave vectors was previously found 
in the Kob-Andersen system by Berthier \cite{Berthier}. Also, 
the increase of $D q^2 \tau_{\alpha}(q)$ has been predicted within the theoretical approach
of Schweizer and Saltzman \cite{Ken}. Finally, Berthier, Chandler and Garrahan
found a similar behavior in one-dimensional facilitated kinetic Ising models \cite{BCG}. 
Here we emphasize that this behavior correlates with strong deviations
from Fickian diffusion visible in the probability
distribution $P\bm{(}\log_{10}(\delta r);t\bm{)}$.

\section{\label{conclusion} Conclusions}

We have conducted an extensive comparison of the predictions of the
mode-coupling theory to Brownian dynamics simulations.  As has been
previously observed, qualitatively, predictions of the
mode-coupling theory agree well with simulations.  Namely,
the mode-coupling theory accurately predicts the two step 
relaxation observed in the scattering functions, the existence of the 
plateau region observed in the mean square displacement,
and the power law behavior of the self diffusion coefficient and
the $\alpha$ relaxation time.  

The mode-coupling theory overestimates the feedback
mechanism in the memory functions, resulting in a transition temperature
$T_c^{theory} = 0.9515$ much greater than the 
temperature $T_c^{sim} = 0.435$ inferred 
from simulations.  The transition temperature found
in simulations is determined by fitting the diffusion
coefficient and the $\alpha$ relaxation time to power
laws.  It should be noted that this 
temperature is sensitive to the range of temperatures used 
in the power law fits \cite{FlennerSzamel2}.  
While the transition temperatures are vastly different,
we find that the mode-coupling theory gives good quantitative
results of many time dependent quantities if 
they are compared at the same reduced temperature 
$\epsilon = (T-T_c)/T_c$.  

The self intermediate scattering functions 
and the mean square displacement resulting from simulations are well described 
by the mode-coupling theory for reduced temperatures 
greater than 0.08.  
For temperatures close to $T_c$ the mode-coupling theory 
predicts divergence of the self intermediate scattering
function's relaxation time and vanishing of the
self diffusion coefficient.  There is no divergence of the
relaxation time or vanishing of the self
diffusion coefficient in the Brownian dynamics simulations.

The non-Gaussian parameter calculated from the simulations
are quite different than the non-Gaussian parameter 
predicted by the mode-coupling theory.  The mode-coupling theory predicts two peaks
in $\alpha_2(t)$ for reduced temperatures $\epsilon \le 1.2989$
for the A particles, and a shoulder at short times and a peak at
longer times for the B particles.    
There is only one peak in the non-Gaussian parameter 
calculated from the simulation for all temperatures for both the A and B
particles.  Furthermore, the position of the 
second peak for the A particles and the only peak
for the B particles predicted by the mode-coupling theory 
has a different temperature dependence than 
the position of the peak observed in the Brownian dynamics simulations.
The mode-coupling theory predicts that the position of the
second peak roughly follows the $\alpha$ relaxation time
close to $T_c$, while the position of the peak for the
Brownian dynamics simulations increases 
slower with decreasing temperature than the $\alpha$ relaxation
time. Finally, the theory underestimates the height of the 
non-Gaussian parameter peak by almost an order of magnitude.

We calculated the probability of the logarithm of single particle
displacements $P\bm{(}\log_{10}(\delta r);t\bm{)}$ which is sensitive
to hopping-like motion.  At high reduced temperatures, there is
no hopping-like motion evident in the Brownian dynamics simulations
and $P\bm{(}\log_{10}(\delta r);t\bm{)}$ is accurately described by the 
mode-coupling theory.  For low reduced temperatures, there is little
agreement between the mode-coupling theory and the simulations.  
At the lowest reduced temperature
studied in this work, $\epsilon = 0.0115$, the hopping-like
motion is very evident in the Brownian dynamics simulations, but
is not predicted by the mode-coupling theory. We believe that this 
hopping-like motion is responsible for the absence of the 
divergence of the relaxation time or vanishing of the self
diffusion coefficient in the simulations.

For low reduced temperatures, the mode-coupling
theory does not predict the proper wave vector dependence of the
$\alpha$ relaxation time.  The theory predicts
that at all temperatures the product $D q^2 \tau(q)$ 
is one for small wave vectors and it decreases monotonically
with increasing wave vector.
In contrast, at low temperatures Brownian dynamics simulations show 
a peak in the product $D q^2 \tau(q)$.  
  
To summarize, we found that close but not too close to the transition
temperature the mode-coupling theory does predict most time-dependent quantities
reasonably well. At very small reduced temperatures there is a hopping-like 
motion present in the simulations which 
is not accounted for by the standard version of the mode-coupling theory. 
Signatures of the hopping-like motion include the two-peak structure of
the probability of the logarithm of particle
displacements and non-trivial wave vector dependence of the
$\alpha$ relaxation time. 

\section*{Acknowledgments}
We gratefully acknowledge the support of NSF Grant No.~CHE 0111152.

\appendix*
\section{Numerical Routines}

In this section we will outline the numerical routines that were used to 
solve many of the equations given in this paper.  Since the time
dependent quantities were solved over many decades in time, it is not 
possible to solve these equations using normal Gaussian quadrature, and 
special algorithms must be implemented.  While these techniques have been 
outlined in the literature previously \cite{FuchsMCT,Miyazaki}, they have not been described in 
detail.  In this appendix we will describe numerical routines
used to calculate integrals of the form
\begin{equation}
\int \mbox{d}\vec{k}\ F(|\vec{q} - \vec{k}|) G(k)
\label{int}
\end{equation}
when F(k) and G(k) are only known on a grid of equally
spaced wave vectors and
and how to solve equations of the form
\begin{equation}
\dot{F}_q(t) = a F_q(t) + \int_{0}^{t} \mbox{d}u\ M_q(F_q,t-u) \dot{F}_q(u)
\label{timeeq}
\end{equation}
for many decades in time.  The dot denotes a derivative 
with respect to time.  It is important to note that the function
$M_q$ not only depends on time $t$, but also on the functions $F_q(t)$. 

The integrals of the memory function in the mode-coupling theory are all 
of the form given by Eq.~\ref{int}.  To calculate the integral, the
functions are generally only known on a grid of equally spaced wave vectors.  Furthermore
the calculation of these integrals are the most computationally expensive part
of the program, and extrapolation between grid points can increase the
calculation time significantly.  The first step is to introduce the change of
variables $\vec{p} = \vec{k} - \vec{q}/2$, then convert to 
spherical polar coordinates.  This allows integration over one angular variable
and the integral becomes
\begin{equation}
2 \pi \int p^2 \sin \phi\ \mbox{d}p\ \mbox{d}\phi\  
F\left(\left|\vec{p} - \frac{\vec{q}}{2} \right|\right)
G\left(\left|\vec{p} + \frac{\vec{q}}{2} \right|\right).
\end{equation}
Then make another change of variables to 
\begin{eqnarray}
x & = & \left| \vec{p} + \frac{\vec{q}}{2} \right| =  \sqrt{p^2 + q^2/4 + p q \cos \phi}\\
y & = & \left| \vec{p} - \frac{\vec{q}}{2} \right| =  \sqrt{p^2 + q^2/4 - p q \cos \phi}  
\end{eqnarray}
which results in the integral
\begin{equation}
\frac{2 \pi}{q} \int_{0}^{\infty} x \mbox{d}x \int_{\left| x - q \right|}^{x+q} y 
\mbox{d}y F(y) G(x)
\label{finalint}
\end{equation}
which can easily be calculated numerically using any number of quadrature techniques.
There is one technical issue in using Eq.~\ref{finalint}.  As $q \rightarrow 0$ the 
integral over $y$ goes to zero, but the integral itself does not go to zero, 
see \textit{e.g.}~equation \ref{mzero}.  While the effect is negligible for larger 
wave vectors, for small $q$ it is better to expand Eq.~\ref{int} in
a Taylor series around $q=0$.  We found that this had to be done to 
obtain accurate small $q$ values of varies quantities, \textit{e.g.}~the 
non-ergodicity parameter.  In this work the Taylor series approximation was always
used to calculate the integrals of the memory functions for the two 
smallest wave vectors. 

Most of the equations solved in this paper has the form given by
Eq.~\ref{timeeq} and they must be solved for many decades in time.
The basic algorithm is as follows.  First an arbitrary time
interval $\Delta t$ is broken into $4N$ equal segments of size
$\delta t = \Delta t/4N$.  It is assumed
that the value of $F_q(t)$ for the first $2N$ segments is known.
For each future time an equation of the form
\begin{equation}
F_q(t) = a H(F_q,t) + b.
\label{fequation}
\end{equation}
is solved.
When $F_q(t_i)$ has been calculated for all $4N$ times 
the time interval $\Delta t^{\prime} = 2 \Delta t$ is doubled.
The new time integral in divided into $4N$ equal segments
of size $\delta t^{\prime} = 2 \delta t$.  Since
$F(t)$ has been calculated up to $\Delta t^{\prime}/2$, 
a mapping of $\{F_q(t_i)\} \rightarrow \{F_q(t_j)\}$ can be defined where
$t_i = i \delta t$ and $t_j = j \delta t^{\prime}$ and  thus
we know $F_q(t)$ for the first $2N$ segments of the new time
interval, and the procedure is repeated.

First we will describe how to convert Eq.~\ref{timeeq} into
an equation of the form \ref{fequation}.  Break the 
integral in Eq.~\ref{timeeq} into two integrals, then
integrate the integral starting at $t=0$ by parts to get
\begin{eqnarray}
\int_{0}^{t} \mbox{d}u\ M_q(t-u) \dot{F}_q(u) & 
= & M_q(t - t_2)F_q(t_2) - M_q(t)F_q(0) \nonumber \\
& & - \int_{0}^{t_2} \mbox{d}u \dot{M}_q(t-u) F_q(u) \nonumber \\
& & + \int_{t_2}^{t} \mbox{d}u M_q(t-u) \dot{F}_q(u)
\label{brint}
\end{eqnarray}
Next make a change of variables in the second integral 
in Eq.~\ref{brint} to $\tau = t-u$ and
break both integrals in Eq.~\ref{brint}
into integrals of length $\Delta t/2$.  This results 
in the following exact form of the integral in Eq.~\ref{brint},
\begin{eqnarray}
M_q(t-t_2)F_q(t_2) & - & M_q(t)F_q(0) \nonumber \\
& - & \sum_{j=1}^{n_1} \int_{t_{j-1}}^{t_j} \mbox{d}u \dot{M}_q(t-u) F_q(u)\nonumber \\
& - & \sum_{j=1}^{n2} \int_{t_{j-1}}^{t_j} \mbox{d}u M_q(u) \dot{F}_q(t-u).\nonumber \\
\end{eqnarray}
Use the approximation that 
\begin{eqnarray}
\int_{t_{j-1}}^{t_j} \mbox{d}u \dot{A}(u)B(u) & \approx &  
\{A(t_j) - A(t_{j-1})\}\frac{1}{\delta t} \int_{t_{j-1}}^{t_j} \mbox{d}u B(u)\nonumber \\
& \approx & \{A(t_j) - A(t_{j-1})\} \mathcal{I}\left[B(t_j)\right]
\end{eqnarray}
where $\mathcal{I}\left[B(t_j)\right] = (1/2)[B(t_{j-1}) + B(t_{j})]$.
We approximate $\dot{F}_q(t_i)$ by
\begin{equation}
\dot{F}_q(t) \approx \frac{1}{2 \delta t}F_q(t_{i-2}) - \frac{2}{\delta t}F_q(t_{i-1}) 
+ \frac{3}{2 \delta t} F_q(t_i), 
\end{equation}
but other approximations for the derivative can be used.  Put all this 
together to get the equation
\begin{equation}
C_1 F_q(t_i) = C_2 M_q(t_i) + C_3
\end{equation}
where
\begin{eqnarray}
C_1 & = & \frac{3}{2 \delta t} - \mathcal{I}\left[M_q(t_1)\right]  - a\nonumber \\
C_2 & = & \mathcal{I}\left[F_q(t_1)\right] - F_q(0) \nonumber \\
C_3 & = & \frac{2}{\delta t} F_q(t_{i-1}) - \frac{1}{2 \delta t} F_q(t_{i-1}) \nonumber \\
& & + M_q(t_{i - i2}) F_q(t_{i2})  - M_q(t_{i-1}) \nonumber \\
& & - F_q(t_{i-1}) \mathcal{I}\left[M_q(t_{i2})\right]\nonumber \\
& & - \sum_{j=2}^{i2} \left\{ M_q(t_{i-j}) - M_q(t_{i-j+1}) \right\} 
\mathcal{I}\left[F_q(t_j)\right] \nonumber \\
& & - \sum_{j=2}^{i-i2} \left\{ F_q(t_{i-j}) - F_q(t_{i-j+1}) \right\} 
\mathcal{I}\left[M_q(t_j)\right] \nonumber \\
\end{eqnarray}
which can be easily recast into the form of Eq.~\ref{fequation}.  Recall that
$M_q(t_i)$ depends on $F_q(t_i)$.  Equation \ref{fequation} can be solved
using any number of techniques used to find a fixed point of a set of 
equations.  There is one equation of the form \ref{fequation}
for every $q$ vector, and $M_q(t_i)$ depends on each $F_q(t_i)$. 

It remains to describe the mapping from $\Delta t$ to $\Delta t^{\prime} = 2 \Delta t$. 
For $1 \le j \le 2N$,
\begin{eqnarray}
F_q(t_{2i}) & \rightarrow & F_q(t_{j}) \\ 
M_q(t_{2i}) & \rightarrow & M_q(t_j).
\end{eqnarray}  
For $1 \le j \le N$,
\begin{eqnarray} 
0.5 \left\{ \mathcal{I}\left[ F_q(t_{2i}) \right] + \mathcal{I}\left[ F_q(t_{2i-1}) 
\right] \right\} 
& \rightarrow & \mathcal{I}\left[ F_q(t_j) \right] \nonumber \\
& & \\
0.5 \left\{ \mathcal{I}\left[ M_q(t_{2i}) \right] + \mathcal{I}\left[ M_q(t_{2i-1}) 
\right] \right\}
& \rightarrow & \mathcal{I}\left[ M_q(t_j) \right] \nonumber \\
\end{eqnarray}
For $N+1 \le j \le 2N$
\begin{eqnarray}
\frac{1}{6} \left\{ F_q(t_{2i}) + 4F_q(t_{2i-1}) + F_q(t_{2i-2}) \right\} & \rightarrow & 
\mathcal{I}\left[ F_q(t_j) \right] \nonumber \\
& & \\
\frac{1}{6} \left\{ M_q(t_{2i}) + 4M_q(t_{2i-1}) + M_q(t_{2i-2}) \right\} & \rightarrow & 
\mathcal{I}\left[ M_q(t_j) \right].\nonumber \\
\end{eqnarray}
Now there are values for $F,M,\mathcal{I}[F],\mathcal{I}[M]$ from 
$0 \le t \le \Delta t^{\prime}/2$ and they can be calculated for 
$\Delta t^{\prime}/2 < t \le \Delta t^{\prime}$.

\end{document}